\documentstyle[preprint,aps,eqsecnum,floats]{revtex}

\begin{document}

\date{\today}

\draft


\title{Commensurate Scale Relations in
       Quantum Chromodynamics
\thanks{Work partially supported by the Department of
Energy, contract DE--AC03--76SF00515 and contract
DE--FG02--93ER--40762.} }

\author{Stanley J. Brodsky}
\address{
Stanford Linear Accelerator Center \\
Stanford University, Stanford, California 94309}

\author{Hung Jung Lu}
\address{
Department of Physics, University of Maryland\\
College Park, Maryland 20742}

\maketitle

\begin{abstract}
We use the BLM method to show that
perturbatively-calculable
observables in QCD, including the annihilation ratio $R_{e^+e^-},$
the heavy quark potential, and  radiative
corrections to structure function
sum rules can be related to each other without renormalization
scale or scheme ambiguity.
The
commensurate scale relations connecting the effective charges for
observables $A$ and $B$ have the form
$\alpha_A(Q_A) = \alpha_B(Q_B)
 \left(1 + r_{A/B} {\alpha_B\over \pi} +\cdots\right),$
where the
coefficient $r_{A/B}$ is independent of the number of flavors $f$
contributing to coupling constant renormalization. The ratio of
scales $Q_A/Q_B$  is unique
at leading order and
guarantees that the observables $A$ and $B$
pass through new quark thresholds
at the same physical scale.
We also show that the commensurate scales satisfy
the renormalization group transitivity rule
which ensures that
predictions in PQCD are independent of the choice of an intermediate
renormalization scheme $C.$ In particular, scale-fixed predictions can be
made without reference to theoretically-constructed renormalization
schemes such as $\overline{\rm MS}.$
QCD can thus be tested in a new and precise way by
checking that the observables track
both in their relative normalization and in
their commensurate scale dependence.

The generalization of the BLM procedure to higher order
assigns a different renormalization scale for each order
in the perturbative series. The scales are determined by
a systematic resummation of running coupling constant effects.
The application of this procedure to relate known physical
observables in QCD gives  surprisingly simple results.
In particular, we find that
up to light-by-light type corrections,
all terms involving $\zeta_3, \zeta_5$ and $\pi^2$
in the relation between the annihilation
ratio $R_{e^+e^-}$ and the Bjorken sum rule for
polarized electroproduction are
automatically absorbed into the renormalization scales.
The     final series  has simple coefficients which are independent
of color:
$\widehat\alpha_{g_1}(Q)
=
\widehat\alpha_R(Q^*)
-
\widehat\alpha_R^2(Q^{**})
+
\widehat\alpha_R^3(Q^{***}),$
where $\widehat\alpha = ({3 C_F / 4 \pi}) \alpha.$
The coefficients in the commensurate scale relation
can be identified with those obtained in
conformally-invariant gauge theory.

\end{abstract}


\vfill
\newpage

\section{Introduction}

One of the most serious difficulties preventing precise tests
of QCD is the scale ambiguity of its perturbative predictions.
Consider a measurable quantity such as
$\rho = R_{e^+e^-}(s)- 3\Sigma e_q^2$.
The PQCD prediction is of the form
\begin{equation}
\rho = r_0\alpha_s(\mu)
\left[1+r_1(\mu)\, {\alpha_s(\mu)\over\pi}+r_2(\mu)\ {\alpha^2_s(\mu)
\over\pi^2} + \cdots\right] \ .
\end{equation}
Here $\alpha_s(\mu) = g^2_s/4\pi$ is the renormalized coupling
defined in a specific renormalization scheme such as $\overline{\rm MS},$
and $\mu$ is a particular choice of renormalization scale.
Since $\rho$ is a physical quantity, its value must be independent of the
choice of $\mu$ as well as the
choice of renormalization scheme.  Nevertheless,
since we only have truncated PQCD predictions to a given order
$\alpha^N_s$, the predictions do depend on $\mu$.  In the
specific case of $R_{e^+e^-}$,
where we have predictions \cite{GorishnyKataevLarin,SurguladzeSamuel}
through order
$\alpha_s^3$, the sensitivity to $\mu$ has been shown to be less than
10\%\ over a large range of $\ell n\, \mu$\cite{SurguladzeSamuel}.
However, in the case of the
hadronic beauty production cross section
$(d\sigma/dp_T^2 )(\bar pp \rightarrow
B + X)$, which has been computed to next-to-leading order in
$\alpha_s$, the prediction\cite{NasonDawsonEllis} \
for the normalization of the
heavy quark $p_T$ distribution at hadron colliders
ranges over a factor of 4 if one chooses one ``physical
value'' such as $\mu= {1\over 4}\
\sqrt{m_B^2+p^2_T}$ rather than an equally
well motivated choice $\mu = \sqrt{m^2_B+p^2_T}$.

There is, in fact, no consensus on how to estimate the theoretical error
due to the scale ambiguity, what constitutes a reasonable range of
physical values, or indeed how to identify what the central value should
be.  Even worse, if we consider the renormalization scale $\mu$ as
totally arbitrary, the next-to-leading coefficient $r_1(\mu)$ in the
perturbative expansion can take on the value zero or any other value.
Thus it is
difficult to assess the convergence of the truncated
series, and finite-order
analyses cannot be meaningfully compared to experiment.

The $\mu$ dependence of the truncated prediction
$\rho_N$ is often used as a
guide to assess the accuracy of the perturbative prediction, since this
dependence reflects the presence of the uncalculated terms.
However, the scale
dependence of $\rho_N$ only reflects one aspect of the total series.
This point has also been recently emphasized by Maxwell {\it et al.}
\cite{BarclayMaxwellReader}.
For example, consider the orthopositronium
${J^{PC}=1^{--}}$ decay rate computed in
quantum electrodynamics:
$\Gamma = \Gamma_0\left[1-10.3\
(\alpha/\pi)+\cdots\right]. $
The large next-to-leading coefficient, $r_1=10.3$
shows that there is important new physics beyond Born approximation.  The
magnitude of the higher order terms
in the decay rate is not related to the
renormalization scale since the QED
coupling $\alpha$ does not run appreciably
at the momentum transfers associated with  positronium decay.

Thus we have a difficult dilemma: If we take $\mu$ as an unset
parameter in PQCD predictions, then we have no reliable way to assess the
accuracy of the truncated series or the parameters extracted from
comparison with experiment.  If we guess a value for $\mu$ and its range,
then we are left with a prediction without an objective guide to its
theoretical precision.  The problem of the scale ambiguity is compounded
in multi-scale problems where several plausible physical scales enter.

In fact three quite distinct methods to set the renormalization scale in
PQCD have been proposed in the literature:

\begin{enumerate}
\item{\sl Fastest Apparent Convergence} (FAC) \cite{Grunberg}.
This method chooses the renormalization scale $\mu$
so that the next-to-leading order coefficient
vanishes: $r_1(\mu) = 0.$

\item {\sl The Principle of Minimum Sensitivity} (PMS) \cite{Stevenson}.
In this procedure, one argues that the best scale is the one that
minimizes the scale dependence of the truncated prediction $\rho_N,$
since
that is a characteristic property of the entire series.
Thus in this method one
chooses $\mu$ at the stationary point $ d \rho_N/ d \mu = 0 .$

\item {\sl Brodsky-Lepage-Mackenzie} (BLM) \cite{BrodskyLepageMackenzie}.
In the BLM scale-fixing method, the scale is chosen such that the
coefficients $r_i$
are independent of the number of quark flavors $f$
renormalizing the gluon
propagators.  In leading order,
one chooses the scale so that $f$ does not
appear in the next-to-leading
order coefficient. That is, if $ r_1(\mu) = r_{10}(\mu) + r_{11}(\mu)
f,$ where $r_{10}(\mu)$ and $r_{11}(\mu)$ are $f$ independent,
then one chooses the scale $\mu$ given by the condition
$ r_{11}(\mu)=0.$ This prescription ensures that, as in quantum
electrodynamics, vacuum polarization contributions due to fermion pairs
are all incorporated into the coupling $\alpha(\mu)$ rather than
the coefficients. In the case of non-Abelian theory, the BLM method
automatically
resums the corresponding
gluon as well as quark vacuum polarization contributions
since the coupling $\alpha_s$ is a function of $\beta_0 \propto
11-{2\over 3} f.$
\end{enumerate}

\begin{figure}[h]
\vbox to 4 in
{\vss\hbox to \hsize
   {\hss
     {\includegraphics{FigureKramerLampe.eps}
     }\hss
   }
}

\caption[FigKramerLampe]{The scale $\mu/\sqrt{s}$ according to
    the BLM (dashed-dotted), PMS (dashed), FAC (full) and
    $\sqrt{y}$ (dotted) procedures for the three-jet rate
    in $e^+e^-$ annihilation, as computed by
    Kramer and Lampe \cite{KramerLampe}.  Notice the strikingly different
    behavior of the BLM scale from the PMS and FAC scales at low $y$.  In
    particular, the latter two methods predict increasing values of $\mu$ as
    the jet invariant mass ${\cal M} < \sqrt (y s)$ decreases.
    \label{FigKramerLampe} }
\end{figure}

These scale-setting methods can give strikingly different results in
practical applications, For example, Kramer and Lampe
have analyzed \cite{KramerLampe}
the application of the FAC, PMS, and BLM methods for
the prediction of jet production fractions in $e^+e^-$ annihilation in
PQCD.  Jets are defined  by clustering particles with invariant mass
less than $\sqrt{y s}$, where $y$ is the resolution parameter and
$\sqrt{s}$ is the total center-of-mass energy.  Physically, one expects
the renormalization scale $\mu$ to reflect the invariant mass of the
jets,  that is, $\mu$ should be of order $\sqrt {y s}$.
For example, in the analogous
problem in QED, the maximum virtuality of the photon jet which sets the
argument of the running coupling $\alpha(Q)$ cannot be
larger than $\sqrt {y s}$.
Thus one expects $\mu $ to decrease as the resolution parameter $y \to
0$.  However, the scales chosen by  the FAC and PMS
methods do not reproduce this
behavior (see Fig. 1):
The predicted scales $\mu_{PMS}(y)$ and $\mu_{FAC}(y)$
rise without bound at small
values for the jet fraction $y.$ On the other hand, the BLM scale has the
correct physical behavior as $y
\to 0$.  Since the argument of the running
coupling  becomes
small using the BLM method,
standard QCD perturbation theory in $\alpha_s[\mu_{BLM}(y)]$
will not be convergent
in the low $y$ domain \cite{BurrowsMasuda}.
In contrast, the scales chosen by
PMS and FAC give no sign that the perturbative
results break down in the soft region.

In this paper we shall use the BLM method to show that all
perturbatively-calculable observables in QCD, including the annihilation
ratio $R_{e^+e^-}(Q^2),$ the heavy quark potential, and the radiative
corrections to the Bjorken sum rule can be related to each other at fixed
relative scales.
The ``commensurate scale relation" for observables $A$ and $B$
in terms of their effective charges  has the form
\begin{equation}
\alpha_A(Q_A) = \alpha_B(Q_B)\ \left(1 + r_{A/B} {\alpha_B\over
\pi} +\cdots\right)\ .
\end{equation}
The ratio of the scales $\lambda_{A/B}=Q_A/Q_B$ is chosen so that
the coefficient $r_{A/B}$ is independent of the
number of flavors $f$ contributing to coupling constant
renormalization. This guarantees
that the effective charges for the
observables $A$ and $B$ pass through new quark thresholds at the
same physical scale.
We shall show that
the value  of $\lambda_{A/B}$
is unique at leading order, and
that the relative scales satisfy the transitivity rule
\cite{BrodskyLuSelfconsistency}
\begin{equation}
\lambda_{A/B} = \lambda_{A/C}~\lambda_{C/B}\ .
\end{equation}
This is equivalent to the group property defined by Peterman and
St\"uckelberg \cite{StueckelbergPeterman}
which ensures that predictions
in PQCD are independent of
the choice of an intermediate renormalization scheme
$C$ \cite{KataevRemark}.
In particular,
scale-fixed predictions can be made without reference to
theoretically-constructed renormalization
schemes such as $\overline{\rm MS};$
QCD can thus be tested by checking that the observables track both
in their relative normalization and commensurate
scale dependence.\cite{Leipzig}

\section{Commensurate Scale Relations}

It is interesting that the task of setting the renormalization scale has
never been considered a problem or ambiguity in perturbative QED.
For example,
the leading-order parallel-helicity
amplitude electron-electron scattering
has the form
\begin{equation}
      {\cal M}_{ee\rightarrow ee}(++;++) =
     {8 \pi s\over t}\ \alpha(t) + {8 \pi s\over u}\ \alpha(u) \ .
\end{equation}
Here
$\alpha(Q) = \alpha(Q_0)/(1 -\Pi[Q^2, Q^2_0, \alpha(Q_0)])$ is the
QED running coupling which sums all vacuum polarization insertions $\Pi$
into the renormalized photon propagator. The value
$\alpha(0)$ is conventionally normalized by Coulomb scattering at
$t=-Q^2=0.$ Notice that both physical scales $t$ and $u$  appear as
arguments of the running coupling constant  in the various
terms contributing
to the scattering cross section; if one chooses
any other scale for the running coupling,
in either the direct or
crossed graph amplitude, then one generates
a spurious geometric series in $f\
(\alpha/\pi)\, \ell n (-t/\mu^2)$ or
$f\ (\alpha/\pi)\, \ell n (-u/\mu^2)$, where $f$
represents the number of fermions contributing
to the vacuum polarization of the photon propagator.

In general, the ``skeleton'' expansion of Feynman amplitudes in QED
guarantees that all dependence of an observable on the variable $f$ is
summed into the running coupling constant; the coefficients
in QED perturbation series are thus always
$f$-independent once the proper
scale in $\alpha$ has been set.  Note that the variable $f$ is defined
to count only vacuum polarization insertions, not light-by-light loops,
since such contributions do not contribute to the coupling constant
renormalization in QED.

The use of the running coupling constant $\alpha(Q)$ in QED allows one
to sum in closed form all proper and
improper vacuum polarization insertions to
all orders, thus going well beyond ordinary perturbation theory.  For
example, consider the perturbative series for the lepton
magnetic anomalous moment:
\begin{equation}
 a_\ell = {\alpha(Q^*)\over 2\pi} + r_2{\alpha^2(Q^{**})\over\pi^2}\
 + r_3\ {\alpha^3(Q^{***})\over\pi^3} + \cdots
\end{equation}
the values
$Q^*=e^{-5/4}m_\ell,$ etc., can be determined either by the explicit
insertion of the running coupling into the
integrand of the Feynman amplitude
and the mean value theorem, or equivalently,
by simply requiring that the
coefficients $r_n$ be independent of $f.$ (Light-by-light scattering
contributions are not related to coupling
constant renormalization and thus
enter explicitly in the order $\alpha^3$ coefficient.)
Thus the formula
for the anomalous moment using the running coupling
is form invariant, identical
for each lepton $\ell=
e, \mu, \tau,$ since the dependence on lepton vacuum
polarization insertions is implicitly contained in
the dependence of the running
coupling constant.  These examples are illustrations
of the general principle
that observables such as the anomalous moments can be related to other
observables such as the heavy lepton potential
$V(Q^2)=-4\pi \alpha(Q^2)/Q^2,$
which can be taken as the empirical
definition of the on-shell scheme usually
used to define $\alpha(Q^2).$

The same procedure can easily be adapted \cite{BrodskyLepageMackenzie}
to non-Abelian theories such as QCD.
One of the most useful observables in QCD is  the heavy quark
potential, since it can be computed in lattice gauge theory from a Wilson
loop, and it can  be extracted
phenomenologically from the heavy quarkonium
spectrum.  If the interacting quarks have infinite mass, then all
radiative correction  are associated with the  exchange diagrams,
rather than the vertex corrections.  It is convenient to write the heavy
quark potential as $V(Q^2)=-4\pi C_F \alpha_V(Q)/Q^2.$ This defines
the ``effective charge" $\alpha_V(Q^2)$ where by definition the
``self-scale" $Q^2= -t$ is the momentum transfer squared.
The subscript $V$
indicates that the coupling is defined through the potential.

In fact, any
perturbatively-calculable physical quantity
can be used to define an effective
charge \cite{Grunberg} by incorporating the entire radiative
correction into its definition; for example
\begin{equation}
R_{e^+e^-}(Q^2) \equiv R^0_{e^+e^-}(Q^2)
\left[1+{\alpha_R(Q)\over \pi}\right] \ ,
\end{equation}
where
$R^0$ is the Born result and $Q^2=s=E_{cm}^2$ is the annihilation energy
squared.  An important result is that all effective charges
$\alpha_A(Q)$ satisfy the Gell-Mann-Low renormalization group equation
with the same $\beta_0$ and $\beta_1;$
different schemes or effective charges
only differ through the third and higher coefficients of the
$\beta$ function.  Thus, any effective
charge can be used as a reference running
coupling constant in QCD to define the   renormalization procedure.
More generally, each effective charge or renormalization scheme,
including $\overline{\rm MS}$, is a special case of the universal
coupling function
$\alpha(Q, \beta_n)$ \cite{BrodskyLuStueckelbergPeterman}.  Peterman and
St\"uckelberg have shown \cite{StueckelbergPeterman}
that all
effective charges are related to each other through a set of evolution
equations in the scheme parameters $\beta_n.$
Physical results relating
observables must of course be independent of
the choice of any intermediate
renormalization scheme.

Let us now consider expanding any observable or effective charge
$\alpha_A(Q_A)$ in terms of $\alpha_V:$
\begin{equation}
\alpha_A(Q_A) = \alpha_V(\mu)
\left[1 + (C+D_{VP}~f)\, {\alpha_V\over \pi} +
\cdots\right]\ .
\end{equation}
Since $\alpha_V$  sums all vacuum polarization
contributions by definition, no coefficient in the series expansion in
$\alpha_V$ can depend on $f;$ {\it i.e.}
all vacuum polarization contributions are
already incorporated into the definition of $\alpha_V.$
Thus we must shift the
scale $\mu$ in the argument of $\alpha_V$ to the scale
\cite{BrodskyLepageMackenzie}
$Q_V = e^{3 D_{VP}(\mu)} \mu$:
\begin{equation}
\alpha_A(Q_A) = \alpha_V({Q_V})
\left[1 + r_1^{A/V} \
{\alpha_V\over \pi} + \cdots\right]\ ,
\label{BLMEq}
\end{equation}
where $r_1^{A/V} = C + (33/2)\, D_{VP}$ is the next-to-leading
coefficient in the expansion of the observable $A$ in scheme $V.$ Thus
the relative scale between the two
observables $A$ and $V$, $\lambda_{A/V}=Q_A/Q_V$, is fixed by
the requirement that the coefficients in the expansion in $\alpha_V$
scheme are independent of vacuum polarization
corrections \cite{CelmasterGonsalvesScaleRatio}.
Alternatively, one
can derive the same result
by explicitly integrating the
one loop integrals in the calculation of the observable $A$ using
$\alpha_V(\ell^2)$ in the integrand, where $\ell^2$ is the four-momentum
transferred squared carried by the gluon.
(In practice  one only needs to
compute the mean-value of $\ell n \ell^2
= \ell n\, Q^2_V $\cite{LepageMackenzie}.)  One
can eliminate the $f$ vacuum polarization
dependence that appears in the
higher order coefficients by allowing a new
scale to appear in each order of
perturbation theory. In practice,  often only
the leading order commensurate
scale  is required in
order to test PQCD to good precision.\cite{CelPMSRemark}

We can compute other  observables $B$ and even   effective
charges such as $\alpha_{\overline{\rm MS}}$ as an expansion
in $\alpha_V$ scheme:
\begin{equation}
\alpha_B(Q_B) =
\alpha_V( Q_V)
\left[1 + r_1^{B/V} \ {\alpha_V\over \pi} + \cdots\right]\ ,
\end{equation}
where $Q_V=Q_B/\lambda_{B/V},$ and
again $r_1^{B/V}$ must be independent of vacuum polarization
contributions. We can now substitute and  eliminate $\alpha_V(Q_V):$
\begin{equation}
\alpha_B(Q_B) = \alpha_A(Q_A)
\left[1 + r_1^{B/A} \ {\alpha_A\over \pi} + \cdots\right]\ ,
\label{BLMEqAB}
\end{equation}
where
$Q_B/Q_A = \lambda_{B/A} = {\lambda_{B/V} / \lambda_{A/V}},$ and
$r_1^{B/A}= r_1^{B/V}-r_1^{A/V}.$ Note also the symmetry property
$\lambda_{B/A} \lambda_{A/B} = 1.$
Alternatively, we can compute the commensurate scale
$Q_A=Q_B/\lambda_{B/A}$ directly by requiring  $r_1^{B/A}$ to be
$f$-independent.  The result is in agreement with the transitivity
rule: the BLM procedure for fixing the
commensurate scale ratio between two
observables is independent of the intermediate renormalization scheme.

The scale-fixed relation between the heavy quark
potential effective charge
and $\alpha_{\overline{\rm MS}}$ is
$\alpha_V(Q)=\alpha_{\overline{\rm MS}}\,(e^{-5/6}Q)[1
- 2 (\alpha_{\overline{\rm MS}}/\pi) + \cdots ]$
\cite{BrodskyLepageMackenzie}.
(The one-loop calculation of $\alpha_V$ in $\overline{\rm MS}$
scheme is given in Ref. \cite{HeavyQuarkPotential}.)

The transitivity and symmetry properties of
the commensurate scales  are the
scale transformations of the renormalization ``group"
as originally defined by Peterman and St\"uckelberg
\cite{StueckelbergPeterman}. The predicted
relation between observables must be independent of the order one makes
substitutions; {\it i.e.} the algebraic path one takes
to relate the observables.  It
is important to note that the PMS method, which
fixes the renormalization scale
by finding the point of minimal sensitivity to $\mu,$ does not satisfy
these group properties \cite{BrodskyLuSelfconsistency}.
The results are chaotic
in the sense that the final
scale depends on the path of applying the PMS procedure.
Furthermore, any method
which fixes the scale in QCD must also be applicable
to Abelian theories such as
QED,   since in the limit of small number of colors
$N_C \to 0$
the perturbative coefficients in QCD
coincide with the perturbative coefficients of an Abelian analog
of QCD \cite{HuetSather}.

The commensurate scale relations
given in Eq. (\ref{BLMEqAB})
provide a new way to test QCD:  One can
compare two observables by checking that their effective charges agree
both in normalization and in their scale dependence.
The ratio of commensurate
scales $\lambda_{A/B}$ is fixed uniquely: it ensures
that both observables $A$
and $B$ pass through heavy quark thresholds
at precisely the same physical
point.
Calculations are often  performed
most advantageously in $\overline{\rm MS}$
scheme, but all reference to such theoretically-constructed
schemes may be eliminated when
comparisons are made between observables.  This also avoids the problem
that one need not expand observables in terms of couplings which have
singular or ill-defined functional dependence.

The physical value of the commensurate scale in
$\alpha_V$ scheme reflects the mean virtuality of the exchanged
gluon.  However, in other schemes,
including $\overline{\rm MS},$ the argument
of
the effective charge is displaced from its physical value.  The relative
scale for a number of observables is indicated in
Table I.  For example, the physical scale for the branching ratio
$\Upsilon \to \gamma X$ when expanded in terms of $\alpha_V$ is
$(1/2.77) M_\Upsilon \sim (1/3) M_\Upsilon,$
which reflects the fact that the final state phase space is divided
among three vector systems.
(When one expands in $\overline{\rm MS}$ scheme, the corresponding
scale is  $0.157 M_\Upsilon.$)
Similarly, the  physical scale appropriate to
the hadronic decays of the $\eta_b$ is
$(1/1.67) M_{\eta_b} \sim (1/2) M_{\eta_b}.$

\begin{center}
\setlength{\unitlength}{1in}
\begin{picture}(4,2.8)

\put(1.7,2.5){$\underline{\rm Table \ I}$}

\put(0.5,2.2){Leading Order Commensurate Scale Relations}

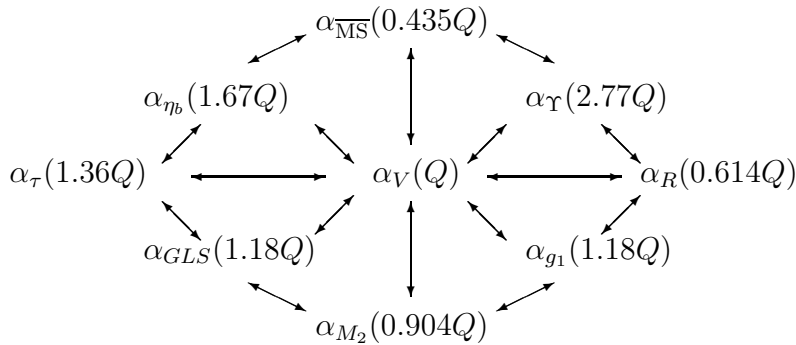
\put(1.6,1.7){$\alpha_{\overline{\rm MS}}(0.435 Q)$}
\put(0.7,1.3){$\alpha_{\eta_b }(1.67 Q)$}
\put(2.7,1.3){$\alpha_\Upsilon (2.77 Q)$}
\put(0.0,0.9){$\alpha_\tau(1.36 Q)$}
\put(1.9,0.9){$\alpha_V(Q)$}
\put(3.3,0.9){$\alpha_R(0.614 Q)$}
\put(0.7,0.5){$\alpha_{GLS}(1.18 Q)$}
\put(2.7,0.5){$\alpha_{g_1}(1.18 Q)$}
\put(1.6,0.1){$\alpha_{M_2}(0.904 Q)$}

\put(1.4,1.6){\vector(2,1){0.15}}
\put(1.4,1.6){\vector(-2,-1){0.15}}
\put(2.7,1.6){\vector(-2,1){0.15}}
\put(2.7,1.6){\vector(2,-1){0.15}}
\put(1.4,0.3){\vector(-2,1){0.15}}
\put(1.4,0.3){\vector(2,-1){0.15}}
\put(2.7,0.3){\vector(2,1){0.15}}
\put(2.7,0.3){\vector(-2,-1){0.15}}
\put(0.8,1.0){\vector(1,1){0.2}}
\put(1.0,1.2){\vector(-1,-1){0.2}}
\put(3.3,1.0){\vector(-1,1){0.2}}
\put(3.1,1.2){\vector(1,-1){0.2}}
\put(1.0,0.6){\vector(-1,1){0.2}}
\put(0.8,0.8){\vector(1,-1){0.2}}
\put(3.1,0.63){\vector(1,1){0.2}}
\put(3.3,0.83){\vector(-1,-1){0.2}}

\put(2.4,1.0){\vector(1,1){0.2}}
\put(2.6,1.2){\vector(-1,-1){0.2}}
\put(1.8,1.0){\vector(-1,1){0.2}}
\put(1.6,1.2){\vector(1,-1){0.2}}
\put(2.6,0.6){\vector(-1,1){0.2}}
\put(2.4,0.8){\vector(1,-1){0.2}}
\put(1.6,0.63){\vector(1,1){0.2}}
\put(1.8,0.83){\vector(-1,-1){0.2}}

\put(2.1,1.35){\vector(0,1){0.25}}
\put(2.1,1.35){\vector(0,-1){0.25}}
\put(2.1,0.55){\vector(0,1){0.25}}
\put(2.1,0.55){\vector(0,-1){0.25}}

\put(1.3,0.93){\vector(1,0){0.35}}
\put(1.3,0.93){\vector(-1,0){0.35}}
\put(2.85,0.93){\vector(1,0){0.35}}
\put(2.85,0.93){\vector(-1,0){0.35}}

\end{picture}
\end{center}

After scale-fixing, the ratio of hadronic to leptonic decay rates for the
$\Upsilon$ has the form \cite{BrodskyLepageMackenzie}
\begin{eqnarray}
{\Gamma(\Upsilon  \to {\rm hadrons})\over
\Gamma(\Upsilon  \to \mu^+ \mu^-)}
&=&
{10 (\pi^2-9) \over 81 \pi e_b^2}\
  {\alpha^3_{\overline{\rm MS}}(0.157 M_\Upsilon)\over \alpha_{\rm QED}^2}
  \left[1 - 14.0(5)\, {\alpha_{\overline{\rm MS}}\over \pi} + \cdots \right]
\\
&=&
{10 (\pi^2-9) \over 81 \pi e_b^2}\
  {\alpha^3_V(0.363 M_\Upsilon)\over \alpha_{\rm QED}^2}
  \left[1 - 8.0(5)\, {\alpha_V\over \pi} + \cdots \right]
{}.
\end{eqnarray}
Thus,
as is the case of positronium decay, the next to leading coefficient
is very large, and  perturbation theory
is not likely to be reliable for this
observable.
On the other hand, the commensurate scales
for the second moment
of the non-singlet structure function $M_2$
and the effective charges in the
Bjorken Sum Rule
(and the Gross-Llewellyn-Smith Sum Rule)  are not far from the physical
value $Q$ when expressed in
$\alpha_V$ scheme.  At large $n$ the commensurate scale for $M_n$ is
proportional
to  $1/\sqrt n$, reflecting the fact  that the  available
phase-space
for parton emission decreases as $n$ increases.
In multiple-scale problems, the
commensurate scale can depend on all of the physical invariants.
For example,
the scale controlling  the evolution equation
for the non-singlet structure
function  depends on $x_{Bj}$ as well as $Q$ \cite{Wong}.
In the case of inclusive reactions which factorize
at leading twist, each structure function, fragmentation
function, and subprocess cross section can have its
own scale.

The commensurate scale relations between
observables can be tested at quite
low momentum transfers, even  where  PQCD relationships would be expected
to break down. It is likely that some of
the higher twist contributions common to
the two observables are also correctly represented
by the commensurate scale relations.   In contrast,
expansions of any observable in
$\alpha_{\overline{\rm MS}}\,(Q)$
must break down at low momentum
transfer since  $\alpha_{\overline{\rm MS}}\,(Q)$
becomes  singular at $Q=\Lambda_{\overline{\rm MS}}.$
(For example, in the 't Hooft scheme where the higher
order $\beta_n=0$ for
$n=2,3,...$ , $\alpha_{\overline{\rm MS}}(Q)$ has a simple pole at
$Q=\Lambda_{\overline{\rm MS}}.$)
The commensurate scale relations allow
tests of QCD without explicit reference
to schemes such as $\overline{\rm MS}.$ It is thus
reasonable to expect that the series
expansions are more convergent when one
relates finite observables to each
other.

\section{Next-to-Leading-Order BLM Formulas}

We will now generalize the BLM procedure
to relate physical observables beyond leading order
\cite{PreviousStudyNLOBLM}. As we
will see, each order in the expansion will acquire
its proper commensurate scale.
Consider the expansion series of a physical effective charge
$\alpha_1(Q)/\pi$ in terms of another physical effective
charge $\alpha_2(Q)/\pi$ \cite{LightByLight}
\begin{eqnarray}
\frac{\alpha_1(Q)}
     {\pi}
&=&
\frac{\alpha_2(Q)}
     {\pi}
+
(A_{12}+B_{12} f)
\left(
       \frac{\alpha_2(Q)}
            {\pi}
\right)^2
\nonumber
\\
& &
+
(C_{12}+D_{12} f+ E_{12} f^2)
\left(
       \frac{\alpha_2(Q)}
            {\pi}
\right)^3
+ \cdots .
\label{Alpha1Alpha2BeforeBLM}
\end{eqnarray}
According to the BLM ansatz,
we can reorganize this series and
resum higher order terms that are induced by
running of the coupling constant effects.
We can perform this procedure order by order,
and absorb these higher order terms
into the renormalization scale of each order.
Postponing the justification, we eventually
obtain a series of the form
\begin{equation}
\frac{\alpha_1(Q)}
     {\pi}
=
\frac{\alpha_2(Q^*)}
     {\pi}
+
{\widetilde A}_{12}
\left(
       \frac{\alpha_2(Q^{**})}
            {\pi}
\right)^2
+
{\widetilde C}_{12}
\left(
       \frac{\alpha_2(Q^{***})}
            {\pi}
\right)^3
+ \cdots .
\label{Alpha1Alpha2AfterBLM}
\end{equation}
Where the running coupling constant effects have
been resummed into the renormalization scales
$Q^*$, $Q^{**}$ and $Q^{***}$.

We now analyze the steps involved
in this process of resummation. First of all,
let us make an important observation:
the $B_{12} f$ term and the $E_{12} f^2$ term
come exclusively from the
one-particle-irreducible vacuum polarization
graphs with one and two fermion loops.
These fermionic contributions belong to the running coupling
constant effects, and should be fully absorbed into the
renormalization scale $Q^*$.

There are also some gluonic
contributions in the running coupling constant effects, and
consequently some part of $A_{12}, C_{12}$ and $D_{12}$
should also be absorbed into $Q^*$. The
exact amount is dictated by
the behavior of the running coupling constant.
The running of $\alpha_2(\mu)$ for a general $SU(N)$ group
can be characterized by
\begin{eqnarray}
\frac{\alpha_2(\mu)}
     {\pi}
&=&
\frac{\alpha_2(\mu_0)}
     {\pi}
- \frac{1}{4}
\left(\frac{11}{3}C_A-\frac{4}{3} T f
\right)
\ell n \left( \frac{\mu^2}
                {\mu_0^2}
    \right)
\left(
       \frac{\alpha_2(\mu_0)}
            {\pi}
\right)^2
\nonumber
\\
& &
+\frac{1}{16}
\Biggl\{
  \left( \frac{11}{3} C_A-\frac{4}{3} T f
  \right)^2
  \ell n^2 \left( \frac{\mu^2}
                    {\mu_0^2}
        \right)
\nonumber
\\
& & \hspace{0.25in}
  -
  \left[ \frac{34}{3} C_A^2-
         \left(\frac{20}{3}C_A+C_F
         \right) T f
  \right]
  \ell n \left( \frac{\mu^2}
                  {\mu_0^2}
      \right)
\Biggr\}
\left(
       \frac{\alpha_2(\mu_0)}
            {\pi}
\right)^3
\nonumber
\\
& &
+\cdots .
\label{RunningAlpha2}
\end{eqnarray}
In the above formula, $C_A=N, C_F=(N^2-1)/2N$
are the quadratic Casimirs of the adjoint and
the fundamental representations. $T$ is the normalization
of the trace of generators of the fundamental
representation: ${\rm Tr}(t^a t^b)= T \delta^{ab}$.
Conventionally $T$ is chosen to be $1/2$.

Let us use $\alpha_2(\mu)$ as the strong coupling
constant in the calculation of the physical effective
charge $\alpha_1(Q)$. If we identify the renormalization
scale $\mu$ with the momentum transfer of the exchanged
gluons in the leading order (LO) Feynman diagrams,
and set $\mu_0=Q$, the
contribution from these diagrams will have the form
\begin{eqnarray}
\left[
\frac{\alpha_1(Q)}
     {\pi}
\right]_{\rm LO}
&=&
\frac{\alpha_2(Q)}
     {\pi}
- \frac{1}{4}
\left( \frac{11}{3} C_A-\frac{4}{3} T f
\right)
L_1
\left(
       \frac{\alpha_2(Q)}
            {\pi}
\right)^2
\nonumber
\\
& &
+ \frac{1}{16}
\Biggl\{
  \left( \frac{11}{3} C_A-\frac{4}{3} T f
  \right)^2
  L_2
\nonumber
\\
& & \hspace{0.25in}
  -
  \left( \frac{34}{3} C_A^2-
         \left(\frac{20}{3}C_A+C_F
         \right) T f
  \right)
  L_1
\Biggr\}
\left(
       \frac{\alpha_2(Q)}
            {\pi}
\right)^3
+ \cdots
\nonumber
\\
&\equiv&
\frac{\alpha_2(Q^*)}
     {\pi} .
\label{LOAlpha1}
\end{eqnarray}
The quantities $L_1$ and $L_2$ are two numerical
constants, which may be interpreted as the mean-values
of the $\ell n(\mu^2/Q^2)$ and $\ell n^2(\mu^2/Q^2)$
terms. In general $L_2 \ne L_1^2$ due to loop
integration or due to the presence of more than one
Feynman diagram involving a tree-level gluon propagator.
This is the structure of terms that we should
absorb in Eq. (\ref{Alpha1Alpha2BeforeBLM}).
In other words, we can shift the renormalization
scale in Eq. (\ref{Alpha1Alpha2BeforeBLM}) until
we fully absorb those higher order terms with
the structure described in Eq. (\ref{LOAlpha1}).
To make this step more clear, let us focus on
the NLO coefficient of Eq. (\ref{Alpha1Alpha2BeforeBLM}),
which is given by $A_{12}+B_{12}f$.
When we perform a scale shift $Q \to Q^*$, we
eliminate the $B_{12} f$ term completely, but at the
same time, we also modify the $A_{12}$ term,
because the net change to the NLO coefficient is
proportional to $\frac{11}{3}C_A - \frac{4}{3} T f$,
as indicated by the structure of the NLO coefficient in
Eq. (\ref{LOAlpha1}). The value of
$L_1$ is determined by the condition of eliminating the
$B_{12} f$ term. Now, proceeding to the NNLO coefficient
of Eq. (\ref{Alpha1Alpha2BeforeBLM}), which is given
by $C_{12} + D_{12} f + E_{12} f^2$, we can determined
the correct value of $L_2$ to eliminate the $E_{12} f^2$
term. The modification to the $C_{12}$ and the $D_{12} f$
term is given by the form of the NNLO coefficient in
Eq. (\ref{LOAlpha1}). The exact form of the scale
shift $Q \to Q^*$ is given later in Eq. (\ref{QStarSUN}).
We should clarify that the scale setting happens
automatically when we use the running coupling
constant inside the Feynman diagram calculation
(like the skeleton expansion in QED). The
mathematical process here of eliminating the $B_{12} f$
and $E_{12}$ terms allows us to determine the
exact amount of the scale shift.

After this first step, our series will look like
\begin{equation}
\frac{\alpha_1(Q)}
     {\pi}
=
\frac{\alpha_2(Q^*)}
     {\pi}
+
{\widetilde A}_{12}
\left(
       \frac{\alpha_2(Q^*)}
            {\pi}
\right)^2
+
\left(
  C'_{12} + D'_{12} f
\right)
\left(
       \frac{\alpha_2(Q^*)}
            {\pi}
\right)^3
+ \cdots .
\label{Alpha1Alpha2HalfwayAfterBLM}
\end{equation}
The next step to follow is now clear. We should
absorb the $D'_{12} f$ term into a new renormalization
scale $Q^{**}$ for the NLO term, since the $f$
dependence comes from the vacuum polarization
corrections to the NLO term. Naturally part of
the $C'_{12}$ term should also be absorbed. From
the form of the running coupling constant in
Eq. (\ref{RunningAlpha2}), we know that the
absorbed term should be proportional to
$\frac{11}{3} C_A - \frac{4}{3} T f$.
After this procedure we arrive to the
form indicated in
Eq. (\ref{Alpha1Alpha2AfterBLM}).
To the order considered here,
we do not have enough information to set the scale $Q^{***}$
of the next-to-next-to-leading order
(NNLO) coupling constant. A sensible choice is
$Q^{***}=Q^{**}$, since this is the renormalization scale
after shifting the scales in the second step of the BLM
procedure.

In practice, most physical observable in perturbative
QCD are computed in the $\overline{\rm MS}$ scheme,
with the running coupling constant fixed at the physical
scale of the process.
Specifically, if the
perturbative series for $\alpha_1(Q)/\pi$ and
$\alpha_2(Q)/\pi$ are
\begin{eqnarray}
\frac{\alpha_1(Q)}
     {\pi}
&=&
\frac{\alpha_{\overline{\rm MS}}(Q)}
     {\pi}
+
(A_1+B_1 f)
\left(
       \frac{\alpha_{\overline{\rm MS}}(Q)}
            {\pi}
\right)^2
\nonumber
\\
& &
+
(C_1+D_1 f+ E_1 f^2)
\left(
       \frac{\alpha_{\overline{\rm MS}}(Q)}
            {\pi}
\right)^3
+ \cdots ,
\label{Alpha1AlphaMS}
\end{eqnarray}
\begin{eqnarray}
\frac{\alpha_2(Q)}
     {\pi}
&=&
\frac{\alpha_{\overline{\rm MS}}(Q)}
     {\pi}
+
(A_2+B_2 f)
\left(
       \frac{\alpha_{\overline{\rm MS}}(Q)}
            {\pi}
\right)^2
\nonumber
\\
& &
+
(C_2+D_2 f+ E_2 f^2)
\left(
       \frac{\alpha_{\overline{\rm MS}}(Q)}
            {\pi}
\right)^3
+ \cdots ,
\label{Alpha2AlphaMS}
\end{eqnarray}
then the coefficients $A_{12}, B_{12}, C_{12}, D_{12}$ and
$E_{12}$ of Eq. (\ref{Alpha1Alpha2BeforeBLM}) are given by
\begin{eqnarray}
A_{12}
&=&
A_1-A_2 ,
\\
B_{12}
&=&
B_1-B_2 ,
\\
C_{12}
&=&
C_1-C_2-2(A_1-A_2)A_2 ,
\\
D_{12}
&=&
D_1-D_2-2(A_1 B_2+A_2 B_1)+4 A_2 B_2 ,
\\
E_{12}
&=&
E_1-E_2
-2(B_1-B_2) B_2 ,
\end{eqnarray}
and the NLO BLM formulas (for a general $SU(N)$ group) are given by
\begin{eqnarray}
\widetilde A_{12}
&=&
A_{12} +
\frac{11}{4}
\frac{C_A}{T} B_{12} ,
\label{A12AfterBLMSUN}
\\
\widetilde C_{12}
&=&
- \frac{3}{16}
  \frac{C_A}
       {T}
  (7C_A+11C_F)
  B_{12}
+ C_{12}
+ \frac{11}{4}
  \frac{C_A}{T}
  D_{12}
+ \frac{121}{16}
  \frac{C_A^2}{T^2}
  E_{12} ,
\label{C12AfterBLMSUN}
\\
Q^*
&=&
Q \exp
\Biggl[ \
   \frac{3}{2 T}
   B_{12}
   + \frac{9}{8 T^2}
   \left( \frac{11}{3} C_A
        - \frac{4}{3} T f
   \right)
   \left(B_{12}^2-E_{12}
   \right)
   \frac{\alpha_2(Q)}
        {\pi} \
\Biggr] ,
\label{QStarSUN}
\\
Q^{**}
&=&
Q \exp
\Biggl\{
\frac{3}{4 T}
\widetilde A_{12}^{-1}
\biggl[ \
    -\frac{1}{4} (5 C_A + 3 C_F) B_{12}
    + D_{12}
    + \frac{11}{2} \frac{C_A}{T} E_{12} \
\biggr] \
\Biggr\} .
\label{QStarStarSUN}
\end{eqnarray}
Notice the presence of $\alpha_2(Q)/\pi$
in the expression of $Q^*$. In general
$Q^*$ will itself be a perturbative series in
$\alpha_2(Q)/\pi$. This fact has first been
pointed out in Ref. \cite{BrodskyLepageMackenzie}, and also
been further explored by Grunberg and Kataev
on the extension of the BLM approach
\cite{GrunbergKataev}. We have exponentiated
the perturbative series, since physically
the renormalization scale $Q^*$ should always
be positive.
To the order considered here,
the scale for the coupling constant in
Eq. (\ref{QStarSUN}) is not well-defined,
but can be chosen to be $Q$. This intrinsic
uncertainty is similar to the $Q^{***}$ scale
uncertainty of the NNLO term in
Eq. (\ref{Alpha1Alpha2AfterBLM}), and
can only be resolved by going to
the next-higher order.

For a general gauge field theory,
it is interesting to point out that the scale
setting procedure described here leads to
the correct expansion series coefficients
in the ``conformal limit"
\cite{DieterMueller}. The conformal
limit is defined by $\beta_0, \beta_1 \to 0$,
and can be reached, for instance, by adding enough spin-half
and scalar quarks. Since all the running
coupling effects have been absorbed into the
renormalization scales, the scale setting method
described here correctly reproduces the expansion
coefficients in this limit. It should be
pointed out that other scale setting methods
in general do not guarantee this feature.

The application of the scale setting formulas to currently
available NNLO QCD quantities gives some very
interesting results. The following is a list of
some effective charges known to the NNLO.

\noindent
1) \underline{$\alpha_R(Q)/\pi$}: the
effective charge obtained in total hadronic
cross section in $e^+e^-$ annihilation, defined by
\cite{GorishnyKataevLarin}
\begin{equation}
R(Q) \equiv 3 \sum_f Q_f^2
\left[ 1+ \frac{\alpha_R(Q)}{\pi}
\right] .
\end{equation}
The perturbative series of $\alpha_R(Q)/\pi$ is
(using $T=1/2$ for the trace normalization)
\begin{eqnarray}
\frac{\alpha_R(Q)}
     {\pi}
&=&
\frac{\alpha_{\rm \overline{MS}}(Q)}
     {\pi}
+
\left(
\frac{\alpha_{\rm \overline{MS}}(Q)}
     {\pi}
\right)^2
\left[
  \left(
        \frac{41}{8}
       -\frac{11}{3} \zeta_3
  \right) C_A
-\frac{1}{8} C_F
+
\left(
- \frac{11}
       {12}
+
\frac{2}
     {3}
\zeta_3
\right) f
\right]
\nonumber
\\
& &+
\left(
\frac{\alpha_{\rm \overline{MS}}(Q)}
     {\pi}
\right)^3
\Biggl\{
  \left(
         \frac{90445}{2592}
        -\frac{2737}{108} \zeta_3
        -\frac{55}{18} \zeta_5
        -\frac{121}{432} \pi^2
  \right)
  C_A^2
\nonumber
\\
& & \hspace{2.5cm}
+ \left(
       -\frac{127}{48}
       -\frac{143}{12} \zeta_3
       +\frac{55}{3} \zeta_5
  \right)
  C_A C_F
-\frac{23}{32} C_F^2
\nonumber
\\
& &\hspace{2.5cm}
+
\biggl[
   \left(
        -\frac{970}{81}
        +\frac{224}{27} \zeta_3
        +\frac{5}{9} \zeta_5
        +\frac{11}{108} \pi^2
   \right)
   C_A
\nonumber
\\
& &\hspace{2.8cm}
+  \left(
        - \frac{29}{96}
        + \frac{19}{6} \zeta_3
        - \frac{10}{3} \zeta_5
   \right)
   C_F
\biggr]
f
\nonumber
\\
& &\hspace{2.5cm}
+
\left(
\frac{151}
     {162}
-
\frac{19}
     {27}
\zeta_3
-
\frac{1}
     {108}
\pi^2
\right)
f^2
\nonumber
\\
& &\hspace{2.5cm}
+
\left(
\frac{11}
     {144}
-
\frac{1}
     {6}
\zeta_3
\right)
\frac{d^{abc}d^{abc}}
     {C_F d(R)}
\frac{\left( \sum_f Q_f
      \right)^2}
     {\sum_f Q_f^2}
\Biggr\} .
\end{eqnarray}
The term containing $(\sum_f Q_f)^2/ \sum_f Q_f^2$
arises from light-by-light diagrams. The
dimension of the quark representation is $d(R)$, which usually
is $N$ for $SU(N)$. For QCD we have $d^{abc} d^{abc} = 40/3 $.
It is interesting to point out that the vector part of the
$Z$ hadronic decay width \cite{LarinRitbergenVermaseren}
shares the same effective charge as $R(Q)$.

\noindent
2) \underline{$\alpha_\tau(M_\tau)/\pi$}: the
effective charge obtained from the perturbative hadronic
decay rate of the $\tau$ lepton, defined by
\cite{GorishnyKataevLarin}
\begin{equation}
R_\tau \equiv 3
\left[ 1+ \frac{\alpha_\tau(M_\tau)}{\pi}
\right] .
\end{equation}
The perturbative series of $\alpha_\tau(M_\tau)/\pi$ is
\begin{eqnarray}
\frac{\alpha_\tau(Q)}
     {\pi}
&=&
\frac{\alpha_{\rm \overline{MS}}(Q)}
     {\pi}
+
\left(
\frac{\alpha_{\rm \overline{MS}}(Q)}
     {\pi}
\right)^2
\left[
  \left(
        \frac{947}{144}
       -\frac{11}{3} \zeta_3
  \right) C_A
-\frac{1}{8} C_F
+
\left(
- \frac{85}
       {72}
+
\frac{2}
     {3}
\zeta_3
\right) f
\right]
\nonumber
\\
& &+
\left(
\frac{\alpha_{\rm \overline{MS}}(Q)}
     {\pi}
\right)^3
\Biggl\{
  \left(
         \frac{559715}{10368}
        -\frac{2591}{72} \zeta_3
        -\frac{55}{18} \zeta_5
        -\frac{121}{432} \pi^2
  \right)
  C_A^2
\nonumber
\\
& & \hspace{2.5cm}
+ \left(
       -\frac{1733}{576}
       -\frac{143}{12} \zeta_3
       +\frac{55}{3} \zeta_5
  \right)
  C_A C_F
-\frac{23}{32} C_F^2
\nonumber
\\
& &\hspace{2.5cm}
+
\biggl[
   \left(
        -\frac{24359}{1296}
        +\frac{73}{6} \zeta_3
        +\frac{5}{9} \zeta_5
        +\frac{11}{108} \pi^2
   \right)
   C_A
\nonumber
\\
& &\hspace{2.8cm}
+  \left(
        - \frac{125}{288}
        + \frac{19}{6} \zeta_3
        - \frac{10}{3} \zeta_5
   \right)
   C_F
\biggr]
f
\nonumber
\\
& &\hspace{2.5cm}
+
\left(
\frac{3935}
     {2592}
-
\frac{19}
     {18}
\zeta_3
-
\frac{1}
     {108}
\pi^2
\right)
f^2
\Biggr\} .
\end{eqnarray}
The appropriate number of flavors to be used here is
$f=3$. There is no light-by-light contributions
in $\alpha_\tau/\pi$.

\noindent
3) \underline{$\alpha_{g_1}(Q)/\pi$}: the
effective charge obtained from the Bjorken sum rule
for polarized electroproduction, defined by
\cite{LarinVermaseren}
\begin{equation}
\int_0^1 d x
\left[
   g_1^{ep}(x,Q^2) - g_1^{en}(x,Q^2)
\right]
\equiv
\frac{1}{3}
\left|
   \frac{g_A}{g_V}
\right|
\left[ 1- \frac{\alpha_{g_1}(Q)}{\pi}
\right] .
\end{equation}
Notice that there are different normalization
conventions in the literature for the polarized
structure functions. Here we follow
the normalization given in Ref. \cite{LarinVermaseren}.
For a recent review on this sum rule, see
Ref. \cite{EllisKarliner}.
The perturbative series of $\alpha_{g_1}(Q)/\pi$ is
\begin{eqnarray}
\frac{\alpha_{g_1}(Q)}
     {\pi}
&=&
\frac{\alpha_{\rm \overline{MS}}(Q)}
     {\pi}
+
\left(
\frac{\alpha_{\rm \overline{MS}}(Q)}
     {\pi}
\right)^2
\left[
\frac{23}{12} C_A - \frac{7}{8} C_F
-
\frac{1}
     {3}
f
\right]
\nonumber
\\
& &+
\left(
\frac{\alpha_{\rm \overline{MS}}(Q)}
     {\pi}
\right)^3
\Biggl\{
\left(
    \frac{5437}{648}
   -\frac{55}{18} \zeta_5
\right) C_A^2
+
\left(
   -\frac{1241}{432}
   +\frac{11}{9} \zeta_3
\right) C_A C_F
+
\frac{1}{32} C_F^2
\nonumber
\\
& &\hspace{2.5cm}
+
\left[
    \left(
         -\frac{3535}{1296}
         -\frac{1}{2} \zeta_3
         +\frac{5}{9} \zeta_5
    \right) C_A
+   \left(
          \frac{133}{864}
         +\frac{5}{18} \zeta_3
    \right) C_F
\right] f
\nonumber
\\
& &\hspace{2.5cm}
+
\frac{115}
     {648}
f^2
\Biggr\} .
\end{eqnarray}
For $\alpha_{g_1}/\pi$ we do not have light-by-light
contributions, either.

\noindent
4) \underline{$\alpha_{F_3}(Q)/\pi$}: the
effective charge obtained from the Gross-Llewellyn Smith sum rule,
defined by
\cite{LarinVermaseren}
\begin{equation}
\int_0^1 d x
\left[
   F_3^{\bar \nu p}(x,Q^2) + F_3^{\nu p}(x,Q^2)
\right]
\equiv
6 \left[ 1- \frac{\alpha_{F_3}(Q)}{\pi}
\right] .
\end{equation}
As pointed out in Ref. \cite{LarinVermaseren},
the effective charge $\alpha_{F_3}/\pi$ differ
from $\alpha_{g_1}/\pi$ only by the light-by-light
contributions. The perturbative series of
$\alpha_{F_3}/\pi$ is given by
the perturbative series of $\alpha_{g_1}/\pi$
plus the additional term in the following formula.
\begin{equation}
\frac{\alpha_{F_3}(Q)}
     {\pi}
=
\frac{\alpha_{g_1}(Q)}
     {\pi}
-
\left(
\frac{\alpha_{\rm \overline{MS}}(Q)}
     {\pi}
\right)^3
\frac{d^{abc} d^{abc}}
     {C_F N_C}
\left(
   -\frac{11}{144}
   +\frac{1}{6} \zeta_3
\right) f .
\end{equation}
For QCD we have $C_F=4/3$, the number of colors
is $N_C=3$, and $d^{abc} d^{abc}=40/3$.
Since the
radiative corrections to the Bjorken sum rule
are identical to those of the
Gross-Llewellyn-Smith sum rule---up to small
corrections of order $\alpha^3_s(Q^2)$,
a basic test of QCD can be made by considering the ratio of the
Gross-Llewellyn-Smith and Bjorken sum rules:
\begin{equation}
R_{GLLS/Bj}(Q^2,\epsilon) =  {
\frac{1}{6}
\int^1_\epsilon dx
\left[ F^{\nu p}_3(x,Q^2)+F^{\bar \nu p}_3(x,Q^2) \right]\over
3 \left|  \frac{g_V}{g_A}
  \right|
\int^1_\epsilon dx \left[ g_1^p(x,Q^2)-g_1^n(x,Q^2) \right]}.
\end{equation}
Since the Regge behavior of the two sum rules is similar, the empirical
extrapolation to $\epsilon \rightarrow 0$ should be relatively free of
systematic error. Moreover,
PQCD predicts
\begin{equation}
 R_{GLLS/Bj}(Q^2, \epsilon \to 0) = 1
+ {\cal O}\left(\alpha^3_s(Q)\right) +
{\cal O}\left( \Lambda_{QCD}^2\over Q^2\right) \ ,
\end{equation}
{\it i.e.}, hard relativistic
corrections to the ratio of the sum rules only enter at
three loops. Thus measurements of the {\it ratio} of the sum rules
could provide a remarkably complication-free test of QCD -
any significant deviation from
$R_{GLLS/Bj}(Q^2,\epsilon \to 0)=1$
must be due to higher twist effects which
should vanish rapidly with increasing $Q^2.$

\noindent
5) \underline{$\alpha_{F_1}(Q)/\pi$}: the
effective charge obtained from the Bjorken sum rule
for deep-inelastic neutrino-nucleon scattering, defined by
\cite{LarinTkachovVermaseren}
\begin{equation}
\int_0^1 d x
\left[
   F_1^{\bar\nu p}(x,Q^2) - F_1^{\nu p}(x,Q^2)
\right]
\equiv
1
-
\frac{1}{2} C_F
\left(
\frac{\alpha_{F_1}(Q)}{\pi}
\right) .
\end{equation}
The perturbative series of $\alpha_{F_1}(Q)/\pi$ is
\begin{eqnarray}
\frac{\alpha_{F_1}(Q)}
     {\pi}
&=&
\frac{\alpha_{\rm \overline{MS}}(Q)}
     {\pi}
+
\left(
\frac{\alpha_{\rm \overline{MS}}(Q)}
     {\pi}
\right)^2
\left[
\frac{91}{36} C_A - \frac{11}{8} C_F
-
\frac{4}
     {9}
f
\right]
\nonumber
\\
& &+
\left(
\frac{\alpha_{\rm \overline{MS}}(Q)}
     {\pi}
\right)^3
\Biggl\{
\left(
    \frac{8285}{648}
   +5 \zeta_3
   -10 \zeta_5
\right) C_A^2
+
\left(
   -\frac{2731}{144}
   -\frac{91}{3} \zeta_3
   +\frac{95}{2} \zeta_5
\right) C_A C_F
\nonumber
\\
& &\hspace{2.5cm}
+
\left(
    \frac{313}{32}
   +\frac{47}{2} \zeta_3
   -35 \zeta_5
\right) C_F^2
\nonumber
\\
& &\hspace{2.5cm}
+
\left[
    \left(
         -\frac{4235}{1296}
         +\frac{7}{6} \zeta_3
         -\frac{5}{3} \zeta_5
    \right) C_A
+   \left(
          \frac{335}{288}
         -\frac{1}{6} \zeta_3
    \right) C_F
\right] f
\nonumber
\\
& &\hspace{2.5cm}
+
\frac{155}
     {648}
f^2
\Biggr\} .
\end{eqnarray}
For $\alpha_{F_1}/\pi$ we do not have light-by-light
contributions, either.

As the first example
of a beyond leading order commensurate scale relation,
we shall express $\alpha_\tau(M_\tau)/\pi$
in terms of $\alpha_R(Q)/\pi$. The appropriate number of
flavors in this case is $f=3$, because $\tau$ decay
occurs below the charm threshold. [Incidentally, the
light-by-light contribution in $\alpha_R(Q)/\pi$ vanishes
for the three flavor case.] The application of the NLO
BLM formulas leads to the following commensurate scale relation
\begin{eqnarray}
\frac{\alpha_\tau(M_\tau)}
     {\pi}
&=&
\frac{\alpha_R(Q^*)}
     {\pi} ,
\label{AlphaTauAlphaRAfterBLM}
\\
Q^*
&=&
M_\tau
\exp
\left[
   - \frac{19}{24}
   - \frac{169}{128}
   \frac{\alpha_R(M_\tau)}
        {\pi}
\right] .
\end{eqnarray}
Notice that all the $\zeta_3, \zeta_5$ and $\pi^2$
terms present in the perturbative series of
$\alpha_R(Q)/\pi$ and $\alpha_\tau(M_\tau)/\pi$
have disappeared when we related these two
physical observables directly. Notice
also the vanishing NLO and NNLO coefficient in
Eq. (\ref{AlphaTauAlphaRAfterBLM}). That is,
up to the NNLO, the two effective charge are
simply related by a BLM scale shift.

As the next example, let us express $\alpha_{g_1}(Q)/\pi$
in terms of $\alpha_R(Q)/\pi$. We will leave the $f$
dependence explicit. The application of the NLO BLM
formulas leads to
\begin{eqnarray}
\frac{\alpha_{g_1}(Q)}{\pi}
&=&
\frac{\alpha_R(Q^*)}{\pi}
-
\frac{3}{4} C_F
\left(
   \frac{\alpha_R(Q^{**})}{\pi}
\right)^2
\nonumber
\\
& &
+
\left[
\frac{9}{16} C_F^2
-
\left(
\frac{11}
     {144}
-
\frac{1}
     {6}
\zeta_3
\right)
\frac{d^{abc}d^{abc}}
     {C_F N}
\frac{\left( \sum_f Q_f
      \right)^2}
     {\sum_f Q_f^2}
\right]
\left(
   \frac{\alpha_R(Q^{***})}{\pi}
\right)^3 ,
\\
Q^*
&=&
Q \exp
\left[
\frac{7}{4}
-
2 \zeta_3
+
\left(
\frac{11}{96}
+\frac{7}{3} \zeta_3
-2 \zeta_3^2
-\frac{\pi^2}{24}
\right)
\left(
   \frac{11}{3} C_A
  -\frac{2}{3} f
\right)
\frac{\alpha_R(Q)}{\pi}
\right],
\\
Q^{**}
&=&
Q \exp
\left[
 \frac{523}{216}
+\frac{28}{9} \zeta_3
-\frac{20}{3} \zeta_5
+
\left(
  -\frac{13}{54}
  +\frac{2}{9} \zeta_3
\right)
\frac{C_A}{C_F}
\right] .
\end{eqnarray}

As explained previously, the scale $Q^{***}$ in the above
expression can be chosen to be $Q^{**}$.
Notice that aside from the light-by-light
contributions, all the $\zeta_3, \zeta_5$ and
$\pi^2$ dependencies have been absorbed into
the renormalization scales $Q^*$ and $Q^{**}$.
Understandably, the $\pi^2$ term should be
absorbed into renormalization scale since it
comes from the analytical continuation of QCD
coupling constant from the spacelike region
to the time-like region.
However, at present we do not have an
full understanding for the disappearance
of the $\zeta_3$ and $\zeta_5$ terms.

For the three flavor case, or neglecting the
light-by-light contribution, the series simplifies to
\begin{equation}
\widehat\alpha_{g_1}(Q)
=
\widehat\alpha_R(Q^*)
-
\widehat\alpha_R^2(Q^{**})
+
\widehat\alpha_R^3(Q^{***}),
\label{AlphaG1AlphaRAfterBLMThreeFlavors}
\end{equation}
where
\begin{eqnarray}
\widehat\alpha_{g_1}(Q)
&=&
\frac{3 C_F}
     {4 \pi}
\alpha_{g_1}(Q),
\\
\widehat\alpha_R(Q)
&=&
\frac{3 C_F}
     {4 \pi}
\alpha_R(Q).
\end{eqnarray}

Observe the asthonishing simplicity of the expansion
series in Eq. (\ref{AlphaG1AlphaRAfterBLMThreeFlavors}).
These last formulas suggest that for the general $SU(N)$ group
the natural expansion parameter is $\widehat\alpha$.
The use of $\widehat\alpha$ also makes explicit
that the same formula is valid for QCD and QED.
That is, in the limit $N_C \to 0$ the perturbative
coefficients in QCD coincide with the perturbative
coefficients of an Abelian analog of QCD \cite{HuetSather}.

Broadhurst and Kataev have recently observed
a number of interesting relations between
$\alpha_R(Q)$ and $\alpha_{g_1}(Q)$
(the ``Seven Wonders") \cite{BroadhurstKataev}.
Our simple result here reinforces
the idea of a ``secret symmetry" between
$\alpha_R(Q)$ and $\alpha_{g_1}(Q)$.
We see that this hidden simplicity is
only revealed after the application
of the NLO BLM scale setting procedure.

The application of the NLO BLM formulas to
related the other effective charges presented
here give results with similar simplicity
(see Appendix),
except for those cases involving $\alpha_{F_1}$.
For instance, the obtained relation for
$\alpha_\tau(M_\tau)$ and $\alpha_{F_1}$ is


\begin{eqnarray}
\frac{\alpha_\tau(M_\tau)}{\pi}
&=&
\frac{\alpha_{F_1}(Q^*)}{\pi}
+
\frac{5}{4} C_F
\left(
   \frac{\alpha_{F_1}(Q^{**})}{\pi}
\right)^2
\nonumber
\\
& &
+
\Biggl\{
      \biggl(
         -\frac{43}{12}
         -\frac{85}{6} \zeta_3
         +\frac{115}{6} \zeta_5
      \biggr) C_A^2
      +
      \biggl(
         10
         + 34 \zeta_3
         - \frac{95}{2} \zeta_5
      \biggr) C_A C_F
\nonumber
\\
& &
      +
      \biggl(
         -\frac{113}{16}
         -\frac{47}{2} \zeta_3
         + 35 \zeta_5
      \biggr) C_F^2
\Biggr\}
\left(
   \frac{\alpha_{F_1}(Q^{***})}{\pi}
\right)^3 ,
\\
Q^*
&=&
M_\tau
\exp
\left[
-
\frac{53}{24}
+
2 \zeta_3
+
\left(
-\frac{143}{384}
-\frac{7}{3} \zeta_3
+2 \zeta_3^2
+\frac{\pi^2}{24}
\right)
\left(
   \frac{11}{3} C_A
  -\frac{2}{3} f
\right)
\frac{\alpha_{F_1}(M_\tau)}{\pi}
\right] ,
\\
Q^{**}
&=&
M_\tau
\exp
\left[
-\frac{47}{20}
+\frac{28}{5} \zeta_3
-4 \zeta_5
+
\left(
  -\frac{1}{2}
  -\frac{28}{15} \zeta_3
  +\frac{8}{3} \zeta_5
\right)
\frac{C_A}{C_F}
\right] .
\end{eqnarray}

We see that the $\pi^2$ terms is absorbed into the
renormalization scale $Q^*$, but the resulting
coefficients in the expansion series are
not as simple as in the previous
examples. In particular, we observe that the
$\zeta_3$ and $\zeta_5$ terms persist in
the NNLO coefficient, which also has a
$C_A$ dependence.

When the scale setting formulas
are applied to relate physical effective
charges to the somewhat artificial
$\overline{\rm MS}$ scheme coupling constant,
no comparable simplicity is observed. For instance, the
application of the formulas to relate
$\alpha_R(Q)$ to $\alpha_{\overline{\rm MS}}(Q)$
leads to
\begin{eqnarray}
\frac{\alpha_R(Q)}
     {\pi}
&=&
\frac{\alpha_{\overline{\rm MS}}(Q^*)}
     {\pi}
+
\tilde A_{12}
\frac{\alpha_{\overline{\rm MS}}^2(Q^{**})}
     {\pi}
+
\tilde C_{12}
\frac{\alpha_{\overline{\rm MS}}^3(Q^{***})}
     {\pi} ,
\\
\tilde A_{12}
&=&
\frac{C_A}{12}-\frac{C_F}{8} ,
\\
\tilde C_{12}
&=&
\left(
   -\frac{53}{144}
   -\frac{11}{4} \zeta_3
\right) C_A^2
+
\left(
   -\frac{101}{192}
   +\frac{11}{4} \zeta_3
\right) C_A C_F
\nonumber
\\
& &
-\frac{23}{32} C_F^2
+ \frac{d^{abc}d^{abc}}
       {3 C_F d(R)}
\left( \frac{11}{48}
     - \frac{1}{2} \zeta_3
\right)
\frac{\left( \sum_f Q_f
      \right)^2}
     {\sum_f Q_f^2} ,
\\
Q^*
&=&
Q \exp
\Biggl[
 -\frac{11}{4}
 + 2 \zeta_3
\nonumber
\\
& & \hspace{0.5in}
+ \left( -\frac{119}{288}
         -\frac{7}{3} \zeta_3
         + 2 \zeta_3^2
         +\frac{\pi^2}{24}
  \right)
\left( \frac{11}{3} C_A
       -\frac{2}{3} f
\right)
\frac{\alpha_{\overline{\rm MS}}(Q)}
     {\pi}
\Biggr] ,
\\
Q^{**}
&=&
Q \exp
\Biggl[
\frac{ (-166-80\zeta_3+160\zeta_5) C_A
       +(111+768\zeta_3-960\zeta_5) C_F }
     {16 C_A - 24 C_F}
\Biggr] .
\end{eqnarray}
These last expressions clearly do not
exhibit the simplicity of the cases
shown in Eq.  (\ref{AlphaTauAlphaRAfterBLM})
or Eq. (\ref{AlphaG1AlphaRAfterBLMThreeFlavors}).

\section{Conclusion}

The problem of the scale ambiguity of PQCD predictions has plagued
attempts to make reliable and precise tests of the theory.  In this
paper we have shown how this problem can be avoided by focussing
on relations between experimentally-measurable observables.
The conventional $\overline{\rm MS}$ renormalization scheme
serves simply as an intermediary between observables.
For example, consider the entire radiative corrections to the
annihilation cross section expressed as the effective charge
$\alpha_R(Q)$ where $Q=\sqrt s$:
\begin{equation}
 R(Q) \equiv 3 \sum_f Q_f^2 \left[ 1+
{\alpha_R(Q) \over \pi} \right].
\end{equation}
Similarly, we can define the entire
radiative correction to the Bjorken sum rule as the effective charge
$\alpha_{g_1}(Q)$ where $Q$ is the lepton momentum transfer:
\begin{equation}
\int_0^1
d x \left[
   g_1^{ep}(x,Q^2) - g_1^{en}(x,Q^2) \right]
   \equiv {1\over 3} \left|g_A \over g_V \right|
   \left[ 1- {\alpha_{g_1}(Q) \over \pi} \right] .
\end{equation}
We now use the
known expressions to three loops in $\overline{\rm MS}$ scheme and choose
the scales $Q^*$ and $Q^{**}$ to re-sum all quark and gluon vacuum
polarization corrections into the running couplings.  The value of these
scales are the physical values which
insure that each observable passes through the heavy quark thresholds at
their respective commensurate physical scales.
The final result is remarkably
simple:
\begin{equation}
{\alpha_{g_1}(Q) \over \pi} = {\alpha_R(Q^*) \over \pi} -
\left( {\alpha_R(Q^{**}) \over \pi} \right)^2 + \left( {\alpha_R(Q^{***})
\over \pi} \right)^3 + . . .
\end{equation}
A fundamental test of QCD is to verify
empirically that the observables track in both normalization and shape as
given by these relations.  The coefficients in the series (aside for a
factor of $C_F$ which can be absorbed in the definition of $\alpha_s$)
are actually independent of color and are the same in Abelian,
non-Abelian, and conformal gauge theory.  The non-Abelian structure of
the theory is reflected in the scales $Q^*$ and $Q^{**}.$ The
commensurate scale relations thus provide fundamental tests of QCD which
can be made increasingly precise and independent of any scheme or
other theoretical convention.

We have also presented in this paper a number of other commensurate
scale relations using the extension of the BLM
method to the next-to-leading order.
We have shown that in each case the application of
the NLO BLM formulas to relate known
physical observables in QCD leads to
results with surprising elegance and simplicity.

In principle, commensurate scale relations allow tests of perturbative
QCD with higher and higher precision as the perturbative expansion grows.
They also provide a new way to specify QCD phenomenology.
Because they relate observables, the commensurate scale relations are
convention-independent; {\it i.e.}, independent of the
normalization conventions
used to define the color $SU(N)$ matrices, etc.
Since the ambiguities
due to scale and scheme choice have been eliminated, one can ask
fundamental questions concerning the nature of the QCD perturbative
expansions, {\it e.g.}, whether the series is convergent or
asymptotic due to
renormalons, etc.\cite{Mueller}.  The precision of the leading twist
series will also allow sensitive tests for higher twist contributions to
physical observables.

We emphasize that any consistent renormalization scheme, $\overline
{\rm MS}$, ${\rm MS}$, ${\rm MOM}$, etc., with any
arbitrary choice of renormalization
scale $\mu,$ can be used in the intermediate stages of analysis.  The
final result, the commensurate scale relation between observables, is
guaranteed to be independent of the choice of intermediate
renormalization scheme since the BLM procedure satisfies the generalized
renormalization group properties of Peterman and Stuckelberg.  An
important computational advantage is that one only needs to compute the
$f$-dependence of the higher order terms in order to specify the lower
order scales in the commensurate scale relations.  In many cases, the
series coefficients in the commensurate scale relations can be determined
from the corresponding Abelian theory; i.e. $N_C \to 0.$

The BLM method and the commensurate scale
relations presented here
can be applied to the whole range of QCD and
standard model processes, making the
tests of theory much more sensitive. The method should also improve
precision tests of electroweak, supersymmetry  and other non-Abelian
theories.
One of the most interesting and important areas of application of
commensurate scale relations will be to the hadronic corrections to
exclusive and inclusive weak decays of heavy quark systems, since the
scale ambiguity in the QCD radiative corrections is at present often the
largest component in the theoretical error entering electroweak
phenomenology.

The commensurate scale relations for some of the
observables discussed in this paper ($\alpha_R,
\alpha_\tau, \alpha_{g_1}$ and $\alpha_{F_3}$)
are universal in the sense that the coefficients of $\widehat \alpha_s$
are independent of color; in fact, they are the same as those for Abelian
gauge theory.  Thus much information on the
structure of the non-Abelian commensurate
scale relations can be obtained
from much simpler Abelian analogs. In fact, in the examples we have
discussed here,
the non-Abelian nature of gauge theory is reflected in the
$\beta$-function coefficients and the choice of second-order scale
$Q^{**}.$   The lack of convergence of
the non-Abelian theory such as renormalon behavior could
show up as a progressive decrease of the higher order
commensurate scales.
The coefficients in these commensurate
scale relations are simply $\pm 1$ and $0$ and
suggest that the underlying relation between observables without
light-by-light contributions is possibly a geometric series.  Note that
the relative correction due to the three loop corrections is $\pm
(\alpha_s/\pi)^2$ or less than $1\%$ for $\alpha_s < 0.3$.

A natural procedure for developing a precision QCD phenomenology is to
choose one effective charge as the canonical definition of the QCD
coupling, and then predict all other observables in terms of this
canonical measure.  Ideally, the heavy quark effective charge
$\alpha_V(Q^2)$ could serve this central role since it can be determined
from both the quarkonium spectrum and from lattice gauge theory.
However, in order for this effective charge to be useful in practice, it
will be necessary to compute the relation of the heavy quark potential to
other schemes through three loops.  At present, the most precisely
theoretically and empirically known effective coupling is
$\alpha_R(Q^2),$ as determined from the annihilation cross section; thus
it is natural to use it as the standard definition.

Alternatively, one can follow historical convention and continue to use
the $\overline {\rm MS}$ scheme as an intermediary between observables.
For definiteness, we can define the $\overline{\rm MS}$
scheme as having all
$\beta_n = 0$ beyond $n=1.$ The commensurate scale relations such as Eq.
(\ref{QStarSUN}), (\ref{QStarStarSUN})
then unambiguously specify all of the scales $Q^*, Q^{**}$, etc.
required to relate $\alpha_{\overline{\rm MS}}$ to the observables.  The
intrinsic QCD scale will then be unambiguously encoded as
$\Lambda_{\overline{\rm MS}}$.  However, there is am intrinsic disadvantage
in using $\alpha_{\overline{\rm MS}}(Q)$ as an expansion parameter: the
function $\alpha_{\overline{\rm MS}}(Q)$ has a simple pole at
$Q=\Lambda_{\overline{\rm MS}}$, whereas observables are by definition
finite.

The BLM scale has also recently been used
by Lepage and Mackenzie \cite{LepageMackenzie}
and their co-workers to improve lattice perturbation theory.
By using the BLM
method one can eliminate $\alpha_{\rm Lattice}$
in favor of $\alpha_V$ thus
avoiding an expansion with artificially large coefficients. The lattice
determination, together with the empirical constraints from the heavy
quarkonium spectra, promises to provide
a well-determined effective charge
$\alpha_V(Q)$ which could be adopted as the QCD standard coupling.

After one fixes the renormalization scale $\mu$
to the BLM value, it is still
useful to compute the logarithmic derivative
of the truncated perturbative prediction
$d \ell n\, \rho_N/
d \ell n \mu$ at the BLM-determined scale.
If this derivative is large, or
equivalently, if the BLM and  PMS scales
strongly differ, then one knows that the
truncated perturbative expansion  cannot be numerically reliable,
since the entire series is independent of $\mu.$
Note that this is a necessary
condition for a  reliable series, not a sufficient
one, as evidenced by the large
coefficients in the positronium and
quarkonium decay widths which appear when
the scales are set correctly. In the case of the two and three
jet decay fractions in $e^+ e^-$ annihilation,
the BLM and PMS scales
strongly differ at low values of the
jet discriminant $y.$ Thus, by using this
criterion, we  establish that the perturbation theory must
fail in the small $y$ regime, requiring
careful resummation of the $\alpha_s
\ell n\, y$ series.
(A more detailed discussion of the sensitivity
of the jet fractions to scale choice and
jet clustering schemes is given in
Ref. \cite{BurrowsMasuda}.)

However, if we restrict the analysis to  jets with
invariant mass
${\cal M} < \sqrt{y s},$ with $0.14 > y > 0.05$, then
we have an ideal
situation, since both the PMS and FAC scales nearly coincide with the
BLM scale
when one computes  jet ratios in the $\overline{\rm MS}$
scheme (See Fig. 1.) {\it i.e.}, the renormalization
scale dependence in this case is minimal
at the BLM scale, and the computed
NLO (next-to-leading order) coefficient is nearly zero.
In fact,  Kramer and Lampe \cite{KramerLampe}
find that the BLM scale
and the NLO PQCD predictions give a consistent description
of the LEP 2-jet and 3-jet
data for $0.14 > y > 0.05$ at the $Z.$ Neglecting possible
uncertainties due to hadronization effects,
this allows a  determination of $\alpha_s$
with remarkably small
error:\cite{KramerLampe}
$\alpha_{\overline{\rm MS}} (M_z)=0.107\pm 0.003,$ which
corresponds to $\Lambda^{(5)}_{\overline{\rm MS}} = 100 \pm 20 $ MeV.

The central principle we have used in our analysis is the fact that
vacuum polarization contributions are summed by the running coupling
constant in gauge theory.  The argument of the running coupling constant
is then fixed by the requirement that all fermion vacuum polarization is
resummed into $\alpha_s$, rather than appear in the coefficients.  The
fact that NLO correction to the scale $Q^*$ is proportional to $\beta_1$
is consistent with the Peterman-St\"uckelberg renormalization group
analysis and is crucial for applying this method to higher order.
We have also seen that this scale-setting procedure leads to
correct expansion coefficients in the conformal limit, since
the beta-function dependence has been resummed into the
renormalization scales.
The same procedure can be applied to multi-scale problems;
in general, the
commensurate scales $Q^*, Q^{**}$, etc. will depend on all of the
available scales.

\section*{ACKNOWLEDGEMENTS}

We wish to thank
G. Bodwin, P. Burrows, L. Dixon,
G. Grunberg,  P. Huet,  G. Ingelman, A. Kataev,
G. Kramer,  B. Lampe, G. P. Lepage,  P. Mackenzie, H. Masuda,
D. M\"uller, M.~Peskin, M. Samuel,
E. Sather, W. K. Wong, and P. Zerwas for helpful conversations.
This work is supported in part by the Department of
Energy, contract DE--AC03--76SF00515 and contract
DE--FG02--93ER--40762.

\appendix
\section*{}

In this Appendix we present the results of applying
the next-to-leading order BLM procedure to relate
the five known effective charges
$\alpha_R(Q), \alpha_\tau(Q), \alpha_{g_1}(Q)$
$\alpha_{F_3}(Q)$ and $\alpha_{F_1}(Q)$.
Notice that in principle $\alpha_\tau(Q)$
is meaningful only for the three-flavor case and for $Q=M_\tau$.
However, here we will use its analytical expression and leave
$Q$ and $f$ dependence explicit. We will leave
the scale $Q^{***}$ in the following formulas unspecified.
This scale in general can be chosen to be $Q^{**}$, or
in the absence of the NLO term, $Q^*$. We use
${\rm Tr}(t^a t^b) = \frac{1}{2} \delta^{ab}$
for the normalization of the trace generators
in this Appendix.


\noindent
\underline{$\alpha_R(Q)$ in terms of $\alpha_\tau(Q)$.}

\begin{eqnarray}
\frac{\alpha_R(Q)}{\pi}
&=&
\frac{\alpha_\tau(Q^*)}{\pi}
+
\left(
\frac{11}
     {144}
-
\frac{1}
     {6}
\zeta_3
\right)
\frac{d^{abc}d^{abc}}
     {C_F N}
\frac{\left( \sum_f Q_f
      \right)^2}
     {\sum_f Q_f^2}
\left(
   \frac{\alpha_\tau(Q^{***})}{\pi}
\right)^3
\\
\ell n(Q^*/Q)
&=&
\frac{19}{24}
+
\frac{169}{1152}
\left(
\frac{11}{3} C_A
-
\frac{2}{3} f
\right)
\frac{\alpha_\tau(Q)}{\pi} .
\end{eqnarray}

\newpage


\noindent
\underline{$\alpha_R(Q)$ in terms of $\alpha_{g_1}(Q)$.}

\begin{eqnarray}
\frac{\alpha_R(Q)}{\pi}
&=&
\frac{\alpha_{g_1}(Q^*)}{\pi}
+
\frac{3}{4} C_F
\left(
   \frac{\alpha_{g_1}(Q^{**})}{\pi}
\right)^2
\nonumber
\\
& &
+
\left[
\frac{9}{16} C_F^2
+
\left(
\frac{11}
     {144}
-
\frac{1}
     {6}
\zeta_3
\right)
\frac{d^{abc}d^{abc}}
     {C_F N}
\frac{\left( \sum_f Q_f
      \right)^2}
     {\sum_f Q_f^2}
\right]
\left(
   \frac{\alpha_{g_1}(Q^{***})}{\pi}
\right)^3 ,
\\
\ell n(Q^*/Q)
&=&
-
\frac{7}{4}
+
2 \zeta_3
+
\left(
-\frac{11}{96}
-\frac{7}{3} \zeta_3
+2 \zeta_3^2
+\frac{\pi^2}{24}
\right)
\left(
   \frac{11}{3} C_A
  -\frac{2}{3} f
\right)
\frac{\alpha_{g_1}(Q)}{\pi} ,
\\
\ell n(Q^{**}/Q)
&=&
-\frac{233}{216}
+\frac{64}{9} \zeta_3
-\frac{20}{3} \zeta_5
+
\left(
  -\frac{13}{54}
  +\frac{2}{9} \zeta_3
\right)
\frac{C_A}{C_F} .
\end{eqnarray}


\noindent
\underline{$\alpha_R(Q)$ in terms of $\alpha_{F_3}(Q)$.}

\begin{eqnarray}
\frac{\alpha_R(Q)}{\pi}
&=&
\frac{\alpha_{F_3}(Q^*)}{\pi}
+
\frac{3}{4} C_F
\left(
   \frac{\alpha_{F_3}(Q^{**})}{\pi}
\right)^2
\nonumber
\\
& &
+
\left\{
\frac{9}{16} C_F^2
+
\left(
\frac{11}
     {144}
-
\frac{1}
     {6}
\zeta_3
\right)
\frac{d^{abc}d^{abc}}
     {C_F N}
\left[
\frac{\left( \sum_f Q_f
      \right)^2}
     {\sum_f Q_f^2}
-f
\right]
\right\}
\left(
   \frac{\alpha_{F_3}(Q^{***})}{\pi}
\right)^3 ,
\\
\ell n(Q^*/Q)
&=&
-
\frac{7}{4}
+
2 \zeta_3
+
\left(
-\frac{11}{96}
-\frac{7}{3} \zeta_3
+2 \zeta_3^2
+\frac{\pi^2}{24}
\right)
\left(
   \frac{11}{3} C_A
  -\frac{2}{3} f
\right)
\frac{\alpha_{F_3}(Q)}{\pi} ,
\\
\ell n(Q^{**}/Q)
&=&
-\frac{233}{216}
+\frac{64}{9} \zeta_3
-\frac{20}{3} \zeta_5
+
\left(
  -\frac{13}{54}
  +\frac{2}{9} \zeta_3
\right)
\frac{C_A}{C_F} .
\end{eqnarray}


\noindent
\underline{$\alpha_R(Q)$ in terms of $\alpha_{F_1}(Q)$.}

\begin{eqnarray}
\frac{\alpha_R(Q)}{\pi}
&=&
\frac{\alpha_{F_1}(Q^*)}{\pi}
+
\frac{5}{4} C_F
\left(
   \frac{\alpha_{F_1}(Q^{**})}{\pi}
\right)^2
+
\Biggl\{
      \biggl(
         -\frac{43}{12}
         -\frac{85}{6} \zeta_3
         +\frac{115}{6} \zeta_5
      \biggr) C_A^2
\nonumber
\\
& &
      +
      \biggl(
         10
         + 34 \zeta_3
         - \frac{95}{2} \zeta_5
      \biggr) C_A C_F
      +
      \biggl(
         -\frac{113}{16}
         -\frac{47}{2} \zeta_3
         + 35 \zeta_5
      \biggr) C_F^2
\nonumber
\\
& &
+
\left(
\frac{11}
     {144}
-
\frac{1}
     {6}
\zeta_3
\right)
\frac{d^{abc}d^{abc}}
     {C_F N}
\frac{\left( \sum_f Q_f
      \right)^2}
     {\sum_f Q_f^2}
\Biggr\}
\left(
   \frac{\alpha_{F_1}(Q^{***})}{\pi}
\right)^3 ,
\\
\ell n(Q^*/Q)
&=&
-
\frac{17}{12}
+
2 \zeta_3
+
\left(
-\frac{65}{288}
-\frac{7}{3} \zeta_3
+2 \zeta_3^2
+\frac{\pi^2}{24}
\right)
\left(
   \frac{11}{3} C_A
  -\frac{2}{3} f
\right)
\frac{\alpha_{F_1}(Q)}{\pi} ,
\\
\ell n(Q^{**}/Q)
&=&
-\frac{187}{120}
+\frac{28}{5} \zeta_3
-4 \zeta_5
+
\left(
  -\frac{1}{2}
  -\frac{28}{15} \zeta_3
  +\frac{8}{3} \zeta_5
\right)
\frac{C_A}{C_F} .
\end{eqnarray}


\noindent
\underline{$\alpha_\tau(Q)$ in terms of $\alpha_R(Q)$.}

\begin{eqnarray}
\frac{\alpha_\tau(Q)}{\pi}
&=&
\frac{\alpha_R(Q^*)}{\pi}
-
\left(
\frac{11}
     {144}
-
\frac{1}
     {6}
\zeta_3
\right)
\frac{d^{abc}d^{abc}}
     {C_F N}
\frac{\left( \sum_f Q_f
      \right)^2}
     {\sum_f Q_f^2}
\left(
   \frac{\alpha_R(Q^{***})}{\pi}
\right)^3
\\
\ell n(Q^*/Q)
&=&
-\frac{19}{24}
-
\frac{169}{1152}
\left(
\frac{11}{3} C_A
-
\frac{2}{3} f
\right)
\frac{\alpha_R(Q)}{\pi} .
\end{eqnarray}


\noindent
\underline{$\alpha_\tau(Q)$ in terms of $\alpha_{g_1}(Q)$.}

\begin{eqnarray}
\frac{\alpha_\tau(Q)}{\pi}
&=&
\frac{\alpha_{g_1}(Q^*)}{\pi}
+
\frac{3}{4} C_F
\left(
   \frac{\alpha_{g_1}(Q^{**})}{\pi}
\right)^2
+
\frac{9}{16} C_F^2
\left(
   \frac{\alpha_{g_1}(Q^{***})}{\pi}
\right)^3 ,
\\
\ell n(Q^*/Q)
&=&
-
\frac{61}{24}
+
2 \zeta_3
+
\left(
-\frac{301}{1152}
-\frac{7}{3} \zeta_3
+2 \zeta_3^2
+\frac{\pi^2}{24}
\right)
\left(
   \frac{11}{3} C_A
  -\frac{2}{3} f
\right)
\frac{\alpha_{g_1}(Q)}{\pi} ,
\\
\ell n(Q^{**}/Q)
&=&
-\frac{101}{54}
+\frac{64}{9} \zeta_3
-\frac{20}{3} \zeta_5
+
\left(
  -\frac{13}{54}
  +\frac{2}{9} \zeta_3
\right)
\frac{C_A}{C_F} .
\end{eqnarray}


\noindent
\underline{$\alpha_\tau(Q)$ in terms of $\alpha_{F_3}(Q)$.}

\begin{eqnarray}
\frac{\alpha_\tau(Q)}{\pi}
&=&
\frac{\alpha_{F_3}(Q^*)}{\pi}
+
\frac{3}{4} C_F
\left(
   \frac{\alpha_{F_3}(Q^{**})}{\pi}
\right)^2
\nonumber
\\
& &
+
\left\{
\frac{9}{16} C_F^2
-
\left(
\frac{11}
     {144}
-
\frac{1}
     {6}
\zeta_3
\right)
\frac{d^{abc}d^{abc}}
     {C_F N}
f
\right\}
\left(
   \frac{\alpha_{F_3}(Q^{***})}{\pi}
\right)^3 ,
\\
\ell n(Q^*/Q)
&=&
-
\frac{61}{24}
+
2 \zeta_3
+
\left(
-\frac{301}{1152}
-\frac{7}{3} \zeta_3
+2 \zeta_3^2
+\frac{\pi^2}{24}
\right)
\left(
   \frac{11}{3} C_A
  -\frac{2}{3} f
\right)
\frac{\alpha_\tau(Q)}{\pi} ,
\\
\ell n(Q^{**}/Q)
&=&
-\frac{101}{54}
+\frac{64}{9} \zeta_3
-\frac{20}{3} \zeta_5
+
\left(
  -\frac{13}{54}
  +\frac{2}{9} \zeta_3
\right)
\frac{C_A}{C_F} .
\end{eqnarray}


\noindent
\underline{$\alpha_\tau(Q)$ in terms of $\alpha_{F_1}(Q)$.}

\begin{eqnarray}
\frac{\alpha_\tau(Q)}{\pi}
&=&
\frac{\alpha_{F_1}(Q^*)}{\pi}
+
\frac{5}{4} C_F
\left(
   \frac{\alpha_{F_1}(Q^{**})}{\pi}
\right)^2
\nonumber
\\
& &
+
\Biggl\{
      \biggl(
         -\frac{43}{12}
         -\frac{85}{6} \zeta_3
         +\frac{115}{6} \zeta_5
      \biggr) C_A^2
      +
      \biggl(
         10
         + 34 \zeta_3
         - \frac{95}{2} \zeta_5
      \biggr) C_A C_F
\nonumber
\\
& &
      +
      \biggl(
         -\frac{113}{16}
         -\frac{47}{2} \zeta_3
         + 35 \zeta_5
      \biggr) C_F^2
\Biggr\}
\left(
   \frac{\alpha_{F_1}(Q^{***})}{\pi}
\right)^3 ,
\\
\ell n(Q^*/Q)
&=&
-
\frac{53}{24}
+
2 \zeta_3
+
\left(
-\frac{143}{384}
-\frac{7}{3} \zeta_3
+2 \zeta_3^2
+\frac{\pi^2}{24}
\right)
\left(
   \frac{11}{3} C_A
  -\frac{2}{3} f
\right)
\frac{\alpha_{F_1}(Q)}{\pi} ,
\\
\ell n(Q^{**}/Q)
&=&
-\frac{47}{20}
+\frac{28}{5} \zeta_3
-4 \zeta_5
+
\left(
  -\frac{1}{2}
  -\frac{28}{15} \zeta_3
  +\frac{8}{3} \zeta_5
\right)
\frac{C_A}{C_F} .
\end{eqnarray}


\noindent
\underline{$\alpha_{g_1}(Q)$ in terms of $\alpha_R(Q)$.}

\begin{eqnarray}
\frac{\alpha_{g_1}(Q)}{\pi}
&=&
\frac{\alpha_R(Q^*)}{\pi}
-
\frac{3}{4} C_F
\left(
   \frac{\alpha_R(Q^{**})}{\pi}
\right)^2
\nonumber
\\
& &
+
\left[
\frac{9}{16} C_F^2
-
\left(
\frac{11}
     {144}
-
\frac{1}
     {6}
\zeta_3
\right)
\frac{d^{abc}d^{abc}}
     {C_F N}
\frac{\left( \sum_f Q_f
      \right)^2}
     {\sum_f Q_f^2}
\right]
\left(
   \frac{\alpha_R(Q^{***})}{\pi}
\right)^3 ,
\\
\ell n(Q^*/Q)
&=&
\frac{7}{4}
-
2 \zeta_3
+
\left(
\frac{11}{96}
+\frac{7}{3} \zeta_3
-2 \zeta_3^2
-\frac{\pi^2}{24}
\right)
\left(
   \frac{11}{3} C_A
  -\frac{2}{3} f
\right)
\frac{\alpha_R(Q)}{\pi} ,
\\
\ell n(Q^{**}/Q)
&=&
 \frac{523}{216}
+\frac{28}{9} \zeta_3
-\frac{20}{3} \zeta_5
+
\left(
  -\frac{13}{54}
  +\frac{2}{9} \zeta_3
\right)
\frac{C_A}{C_F} .
\end{eqnarray}


\noindent
\underline{$\alpha_{g_1}(Q)$ in terms of $\alpha_\tau(Q)$.}

\begin{eqnarray}
\frac{\alpha_{g_1}(Q)}{\pi}
&=&
\frac{\alpha_\tau(Q^*)}{\pi}
-
\frac{3}{4} C_F
\left(
   \frac{\alpha_\tau(Q^{**})}{\pi}
\right)^2
+
\frac{9}{16} C_F^2
\left(
   \frac{\alpha_\tau(Q^{***})}{\pi}
\right)^3 ,
\\
\ell n(Q^*/Q)
&=&
\frac{61}{24}
-
2 \zeta_3
+
\left(
\frac{301}{1152}
+\frac{7}{3} \zeta_3
-2 \zeta_3^2
-\frac{\pi^2}{24}
\right)
\left(
   \frac{11}{3} C_A
  -\frac{2}{3} f
\right)
\frac{\alpha_\tau(Q)}{\pi} ,
\\
\ell n(Q^{**}/Q)
&=&
 \frac{347}{108}
+\frac{28}{9} \zeta_3
-\frac{20}{3} \zeta_5
+
\left(
  -\frac{13}{54}
  +\frac{2}{9} \zeta_3
\right)
\frac{C_A}{C_F} .
\end{eqnarray}


\noindent
\underline{$\alpha_{g_1}(Q)$ in terms of $\alpha_{F_3}(Q)$.}

\begin{eqnarray}
\frac{\alpha_{g_1}(Q)}{\pi}
&=&
\frac{\alpha_{F_3}(Q^*)}{\pi}
-
\left(
\frac{11}
     {144}
-
\frac{1}
     {6}
\zeta_3
\right)
\frac{d^{abc}d^{abc}}
     {C_F N}
f
\left(
   \frac{\alpha_{F_3}(Q^{***})}{\pi}
\right)^3 ,
\\
\ell n(Q^*/Q)
&=&
0.
\end{eqnarray}


\noindent
\underline{$\alpha_{g_1}(Q)$ in terms of $\alpha_{F_1}(Q)$.}

\begin{eqnarray}
\frac{\alpha_{g_1}(Q)}{\pi}
&=&
\frac{\alpha_{F_1}(Q^*)}{\pi}
+
\frac{1}{2} C_F
\left(
   \frac{\alpha_{F_1}(Q^{**})}{\pi}
\right)^2
\nonumber
\\
& &
+
\Biggl\{
      \biggl(
         -\frac{43}{12}
         -\frac{85}{6} \zeta_3
         +\frac{115}{6} \zeta_5
      \biggr) C_A^2
      +
      \biggl(
         10
         + 34 \zeta_3
         - \frac{95}{2} \zeta_5
      \biggr) C_A C_F
\nonumber
\\
& &
      +
      \biggl(
         -\frac{67}{8}
         -\frac{47}{2} \zeta_3
         + 35 \zeta_5
      \biggr) C_F^2
\Biggr\}
\left(
   \frac{\alpha_{F_1}(Q^{***})}{\pi}
\right)^3 ,
\\
\ell n(Q^*/Q)
&=&
\frac{1}{3}
-\frac{1}{9}
\left(
   \frac{11}{3} C_A
  -\frac{2}{3} f
\right)
\frac{\alpha_{F_1}(Q)}{\pi} ,
\\
\ell n(Q^{**}/Q)
&=&
-\frac{37}{36}
+\frac{4}{3} \zeta_3
+
\left(
  -\frac{8}{9}
  -5 \zeta_3
  +\frac{20}{3} \zeta_5
\right)
\frac{C_A}{C_F} .
\end{eqnarray}


\noindent
\underline{$\alpha_{F_3}(Q)$ in terms of $\alpha_R(Q)$.}

\begin{eqnarray}
\frac{\alpha_{F_3}(Q)}{\pi}
&=&
\frac{\alpha_R(Q^*)}{\pi}
-
\frac{3}{4} C_F
\left(
   \frac{\alpha_R(Q^{**})}{\pi}
\right)^2
\nonumber
\\
& &
+
\left\{
\frac{9}{16} C_F^2
-
\left(
\frac{11}
     {144}
-
\frac{1}
     {6}
\zeta_3
\right)
\frac{d^{abc}d^{abc}}
     {C_F N}
\left[
\frac{\left( \sum_f Q_f
      \right)^2}
     {\sum_f Q_f^2}
-f
\right]
\right\}
\left(
   \frac{\alpha_R(Q^{***})}{\pi}
\right)^3 ,
\\
\ell n(Q^*/Q)
&=&
\frac{7}{4}
-
2 \zeta_3
+
\left(
\frac{11}{96}
+\frac{7}{3} \zeta_3
-2 \zeta_3^2
-\frac{\pi^2}{24}
\right)
\left(
   \frac{11}{3} C_A
  -\frac{2}{3} f
\right)
\frac{\alpha_R(Q)}{\pi} ,
\\
\ell n(Q^{**}/Q)
&=&
 \frac{523}{216}
+\frac{28}{9} \zeta_3
-\frac{20}{3} \zeta_5
+
\left(
  -\frac{13}{54}
  +\frac{2}{9} \zeta_3
\right)
\frac{C_A}{C_F} .
\end{eqnarray}

\newpage


\noindent
\underline{$\alpha_{F_3}(Q)$ in terms of $\alpha_\tau(Q)$.}

\begin{eqnarray}
\frac{\alpha_{F_3}(Q)}{\pi}
&=&
\frac{\alpha_\tau(Q^*)}{\pi}
-
\frac{3}{4} C_F
\left(
   \frac{\alpha_\tau(Q^{**})}{\pi}
\right)^2
\nonumber
\\
& &
+
\left\{
\frac{9}{16} C_F^2
+
\left(
\frac{11}
     {144}
-
\frac{1}
     {6}
\zeta_3
\right)
\frac{d^{abc}d^{abc}}
     {C_F N}
f
\right\}
\left(
   \frac{\alpha_\tau(Q^{***})}{\pi}
\right)^3 ,
\\
\ell n(Q^*/Q)
&=&
\frac{61}{24}
-
2 \zeta_3
+
\left(
\frac{301}{1152}
+\frac{7}{3} \zeta_3
-2 \zeta_3^2
-\frac{\pi^2}{24}
\right)
\left(
   \frac{11}{3} C_A
  -\frac{2}{3} f
\right)
\frac{\alpha_\tau(Q)}{\pi} ,
\\
\ell n(Q^{**}/Q)
&=&
 \frac{347}{108}
+\frac{28}{9} \zeta_3
-\frac{20}{3} \zeta_5
+
\left(
  -\frac{13}{54}
  +\frac{2}{9} \zeta_3
\right)
\frac{C_A}{C_F} .
\end{eqnarray}


\noindent
\underline{$\alpha_{F_3}(Q)$ in terms of $\alpha_{g_1}(Q)$.}

\begin{eqnarray}
\frac{\alpha_{F_3}(Q)}{\pi}
&=&
\frac{\alpha_{g_1}(Q^*)}{\pi}
+
\left(
\frac{11}
     {144}
-
\frac{1}
     {6}
\zeta_3
\right)
\frac{d^{abc}d^{abc}}
     {C_F N}
f
\left(
   \frac{\alpha_{g_1}(Q^{***})}{\pi}
\right)^3 ,
\\
\ell n(Q^*/Q)
&=&
0.
\end{eqnarray}


\noindent
\underline{$\alpha_{F_3}(Q)$ in terms of $\alpha_{F_1}(Q)$.}

\begin{eqnarray}
\frac{\alpha_{F_3}(Q)}{\pi}
&=&
\frac{\alpha_{F_1}(Q^*)}{\pi}
+
\frac{1}{2} C_F
\left(
   \frac{\alpha_{F_1}(Q^{**})}{\pi}
\right)^2
+
\Biggl\{
      \biggl(
         -\frac{43}{12}
         -\frac{85}{6} \zeta_3
         +\frac{115}{6} \zeta_5
      \biggr) C_A^2
\nonumber
\\
& &
      +
      \biggl(
         10
         + 34 \zeta_3
         - \frac{95}{2} \zeta_5
      \biggr) C_A C_F
      +
      \biggl(
         -\frac{67}{8}
         -\frac{47}{2} \zeta_3
         + 35 \zeta_5
      \biggr) C_F^2
\nonumber
\\
& &
+
\left(
\frac{11}
     {144}
-
\frac{1}
     {6}
\zeta_3
\right)
\frac{d^{abc}d^{abc}}
     {C_F N}
f
\Biggr\}
\left(
   \frac{\alpha_{F_1}(Q^{***})}{\pi}
\right)^3 ,
\\
\ell n(Q^*/Q)
&=&
\frac{1}{3}
-\frac{1}{9}
\left(
   \frac{11}{3} C_A
  -\frac{2}{3} f
\right)
\frac{\alpha_{F_1}(Q)}{\pi} ,
\\
\ell n(Q^{**}/Q)
&=&
-\frac{37}{36}
+\frac{4}{3} \zeta_3
+
\left(
  -\frac{8}{9}
  -5 \zeta_3
  +\frac{20}{3} \zeta_5
\right)
\frac{C_A}{C_F} .
\end{eqnarray}


\noindent
\underline{$\alpha_{F_1}(Q)$ in terms of $\alpha_R(Q)$.}

\begin{eqnarray}
\frac{\alpha_{F_1}(Q)}{\pi}
&=&
\frac{\alpha_R(Q^*)}{\pi}
-
\frac{5}{4} C_F
\left(
   \frac{\alpha_R(Q^{**})}{\pi}
\right)^2
+
\Biggl\{
      \biggl(
          \frac{43}{12}
         +\frac{85}{6} \zeta_3
         -\frac{115}{6} \zeta_5
      \biggr) C_A^2
\nonumber
\\
& &
      +
      \biggl(
         - 10
         - 34 \zeta_3
         + \frac{95}{2} \zeta_5
      \biggr) C_A C_F
      +
      \biggl(
          \frac{163}{16}
         +\frac{47}{2} \zeta_3
         - 35 \zeta_5
      \biggr) C_F^2
\nonumber
\\
& &
-
\left(
\frac{11}
     {144}
-
\frac{1}
     {6}
\zeta_3
\right)
\frac{d^{abc}d^{abc}}
     {C_F N}
\frac{\left( \sum_f Q_f
      \right)^2}
     {\sum_f Q_f^2}
\Biggr\}
\left(
   \frac{\alpha_R(Q^{***})}{\pi}
\right)^3 ,
\\
\ell n(Q^*/Q)
&=&
\frac{17}{12}
-
2 \zeta_3
+
\left(
\frac{65}{288}
+\frac{7}{3} \zeta_3
-2 \zeta_3^2
-\frac{\pi^2}{24}
\right)
\left(
   \frac{11}{3} C_A
  -\frac{2}{3} f
\right)
\frac{\alpha_R(Q)}{\pi} ,
\\
\ell n(Q^{**}/Q)
&=&
\frac{51}{40}
+\frac{8}{5} \zeta_3
-4 \zeta_5
+
\left(
  -\frac{1}{2}
  -\frac{28}{15} \zeta_3
  +\frac{8}{3} \zeta_5
\right)
\frac{C_A}{C_F} .
\end{eqnarray}


\noindent
\underline{$\alpha_{F_1}(Q)$ in terms of $\alpha_\tau(Q)$.}

\begin{eqnarray}
\frac{\alpha_{F_1}(Q)}{\pi}
&=&
\frac{\alpha_\tau(Q^*)}{\pi}
-
\frac{5}{4} C_F
\left(
   \frac{\alpha_\tau(Q^{**})}{\pi}
\right)^2
\nonumber
\\
& &
+
\Biggl\{
      \biggl(
          \frac{43}{12}
         +\frac{85}{6} \zeta_3
         -\frac{115}{6} \zeta_5
      \biggr) C_A^2
      +
      \biggl(
         - 10
         - 34 \zeta_3
         + \frac{95}{2} \zeta_5
      \biggr) C_A C_F
\nonumber
\\
& &
      +
      \biggl(
          \frac{163}{16}
         +\frac{47}{2} \zeta_3
         - 35 \zeta_5
      \biggr) C_F^2
\Biggr\}
\left(
   \frac{\alpha_\tau(Q^{***})}{\pi}
\right)^3 ,
\\
\ell n(Q^*/Q)
&=&
\frac{53}{24}
-
2 \zeta_3
+
\left(
 \frac{143}{384}
+\frac{7}{3} \zeta_3
-2 \zeta_3^2
-\frac{\pi^2}{24}
\right)
\left(
   \frac{11}{3} C_A
  -\frac{2}{3} f
\right)
\frac{\alpha_\tau(Q)}{\pi} ,
\\
\ell n(Q^{**}/Q)
&=&
\frac{31}{15}
+\frac{8}{5} \zeta_3
-4 \zeta_5
+
\left(
  -\frac{1}{2}
  -\frac{28}{15} \zeta_3
  +\frac{8}{3} \zeta_5
\right)
\frac{C_A}{C_F} .
\end{eqnarray}


\noindent
\underline{$\alpha_{F_1}(Q)$ in terms of $\alpha_{g_1}(Q)$.}

\begin{eqnarray}
\frac{\alpha_{F_1}(Q)}{\pi}
&=&
\frac{\alpha_{g_1}(Q^*)}{\pi}
-
\frac{1}{2} C_F
\left(
   \frac{\alpha_{g_1}(Q^{**})}{\pi}
\right)^2
\nonumber
\\
& &
+
\Biggl\{
      \biggl(
          \frac{43}{12}
         +\frac{85}{6} \zeta_3
         -\frac{115}{6} \zeta_5
      \biggr) C_A^2
      +
      \biggl(
         - 10
         - 34 \zeta_3
         + \frac{95}{2} \zeta_5
      \biggr) C_A C_F
\nonumber
\\
& &
      +
      \biggl(
          \frac{71}{8}
         +\frac{47}{2} \zeta_3
         - 35 \zeta_5
      \biggr) C_F^2
\Biggr\}
\left(
   \frac{\alpha_{g_1}(Q^{***})}{\pi}
\right)^3 ,
\\
\ell n(Q^*/Q)
&=&
-\frac{1}{3}
+\frac{1}{9}
\left(
   \frac{11}{3} C_A
  -\frac{2}{3} f
\right)
\frac{\alpha_{g_1}(Q)}{\pi} ,
\\
\ell n(Q^{**}/Q)
&=&
-\frac{61}{36}
+\frac{4}{3} \zeta_3
+
\left(
  -\frac{8}{9}
  -5 \zeta_3
  +\frac{20}{3} \zeta_5
\right)
\frac{C_A}{C_F} .
\end{eqnarray}


\noindent
\underline{$\alpha_{F_1}(Q)$ in terms of $\alpha_{F_3}(Q)$.}

\begin{eqnarray}
\frac{\alpha_{F_1}(Q)}{\pi}
&=&
\frac{\alpha_{F_3}(Q^*)}{\pi}
-
\frac{1}{2} C_F
\left(
   \frac{\alpha_{F_3}(Q^{**})}{\pi}
\right)^2
+
\Biggl\{
      \biggl(
          \frac{43}{12}
         +\frac{85}{6} \zeta_3
         -\frac{115}{6} \zeta_5
      \biggr) C_A^2
\nonumber
\\
& &
      +
      \biggl(
         - 10
         - 34 \zeta_3
         + \frac{95}{2} \zeta_5
      \biggr) C_A C_F
      +
      \biggl(
          \frac{71}{8}
         +\frac{47}{2} \zeta_3
         - 35 \zeta_5
      \biggr) C_F^2
\nonumber
\\
& &
-
\left(
\frac{11}
     {144}
-
\frac{1}
     {6}
\zeta_3
\right)
\frac{d^{abc}d^{abc}}
     {C_F N}
f
\Biggr\}
\left(
   \frac{\alpha_{F_3}(Q^{***})}{\pi}
\right)^3 ,
\\
\ell n(Q^*/Q)
&=&
-\frac{1}{3}
+\frac{1}{9}
\left(
   \frac{11}{3} C_A
  -\frac{2}{3} f
\right)
\frac{\alpha_{F_3}(Q)}{\pi} ,
\\
\ell n(Q^{**}/Q)
&=&
-\frac{61}{36}
+\frac{4}{3} \zeta_3
+
\left(
  -\frac{8}{9}
  -5 \zeta_3
  +\frac{20}{3} \zeta_5
\right)
\frac{C_A}{C_F} .
\end{eqnarray}



\begin{references}
\bibitem[\dag]{CelmasterGonsalvesScaleRatio}
The commensurate ratio $\lambda_{A/V}=Q_A/Q_V$ differs
from the often-called Celmaster-Gonsalves relation
\cite{BarclayMaxwellReader,Stevenson,CelmasterGonsalves,DharGupta}
for the scales $\Lambda_A/\Lambda_V$. As discussed
in \cite{BrodskyLuStueckelbergPeterman}, a shift in the renormalization
scale as given by $\Lambda_A/\Lambda_V$ leads to
a vanishing NLO coefficient, which is equivalent to
the FAC criterion. In contrast, in the BLM approach, only
the $f$ part vanishes.
\bibitem[\ddag]{CelPMSRemark}
Celmaster and Stevenson
\cite{CelmasterStevenson}
have observed that the NLO coefficient
$r_1^{B/A}$ in an expansion series such as Eq.
(\ref{BLMEq}) depends on the quantity $A$ used in
the right-hand-side, and they have expressed concern
that this fact implies a ``scheme
dependence" of the BLM procedure. This is obviously not
a defect:
when
one expands an observable $B$ in terms of
different observables $A$, one clearly obtains
different coefficients.
\bibitem[**]{PreviousStudyNLOBLM}
The extension of the BLM procedure to higher
orders has also been discussed recently
by Grunberg and Kataev \cite{GrunbergKataev} and by Samuel and
Surguladze \cite{SurguladzeSamuel}.
\bibitem[\dag\dag]{LightByLight}
We must exclude from the analysis the potential $f$ dependence
in the NNLO term induced by light-by-light diagrams.
These diagrams are finite and
do not participate in the renormalization of
the running coupling constant. As a convention,
the coefficient $D_{12}$ in Eq. (\ref{Alpha1Alpha2BeforeBLM})
will include only the $f$ dependence
from the running of the coupling constant. The extra
$f$-dependent terms from light-by-light scattering
diagrams will be considered as part of the
$C_{12}$. This separation
is straightforward in the practical examples
considered in this paper.
\end{references}
\end{document}